\documentclass[a4paper,11pt]{article}
\usepackage{a4wide}
\usepackage[utf8]{inputenc}
\usepackage{graphics,graphicx}
\usepackage{amsmath,amssymb}
\usepackage{xcolor}
\usepackage{physics}
\usepackage{tikz}
\usepackage{pgfplots}
\usepackage{cite}
\usepackage{hyperref}

\usepackage{cleveref}

\numberwithin{equation}{section}

\begin{document}
\thispagestyle{empty}
\begin{flushright}
UWThPh2019-14
\end{flushright}
\vspace{1cm}
\begin{center}
{\LARGE\bf Refined swampland distance conjecture and exotic hybrid Calabi-Yaus}
\end{center}
\vspace{8mm}
\begin{center}
{\large David Erkinger\footnote{{\tt david.josef.erkinger@univie.ac.at}},
  Johanna Knapp\footnote{{\tt johanna.knapp@univie.ac.at}}}
\end{center}
\vspace{3mm}
\begin{center}
{\em Mathematical Physics Group, University of Vienna\\ Boltzmanngasse 5, 1090  Vienna, Austria}
\end{center}
\vspace{15mm}
\begin{abstract}
\noindent We test the refined swampland distance conjecture in the K\"ahler moduli space of exotic one-parameter Calabi-Yaus. We focus on examples with pseudo-hybrid points. These points, whose properties are not well-understood, are at finite distance in the moduli space. We explicitly compute the lengths of geodesics from such points to the large volume regime and show that the refined swampland distance conjecture holds. To compute the metric we use the sphere partition function of the gauged linear sigma model. We discuss several examples in detail, including one example associated to a gauged linear sigma model with non-abelian gauge group. 
\end{abstract}
\newpage
\setcounter{tocdepth}{1}
\tableofcontents
\setcounter{footnote}{0}
\section{Introduction}
\label{sec-introduction}
Initiated by \cite{Vafa:2005ui,Ooguri:2006in}, there has been tremendous activity around the question of how to characterize consistent UV complete theories of quantum gravity, referred to as the landscape, as opposed to the swampland which encompasses all theories that do not have this property. This has led to a whole zoo of swampland conjectures based upon which one can decide whether a given theory of quantum gravity is in the landscape or in the swampland.  A recent review of the state-of-the-art, including hundreds of references, can be found in \cite{Palti:2019pca}.

As string theory is believed to be a consistent theory of quantum gravity, any low-energy effective theory that comes from string theory should be part of the landscape. In the landscape of Calabi-Yau (CY) compactifications, those corners corresponding to large volume regions are the most studied. The inspiration for many conjectures comes from the typical properties of string theory at large volume. Considering regions in the moduli space that are far away from large volume points provides a non-trivial test for the validity of the conjectures. In this work we aim to test one of the conjectures in more exotic corners of the stringy moduli space.  

Concretely, we will test the refined swampland distance conjecture (RSDC) \cite{Baume:2016psm,Klaewer:2016kiy,Palti:2017elp}.  The swampland distance conjecture (SDC) was first proposed in \cite{Ooguri:2006in} and states that at infinite distance from a given point in the moduli space the effective theory breaks down because an infinite tower of light states appears. The information that enters these conjectures can be readily obtained from standard techniques in string theory and has been tested in concrete settings \cite{Grimm:2018ohb,Lee:2018urn,Lee:2018spm,Grimm:2018cpv,Gonzalo:2018guu,Corvilain:2018lgw,Scalisi:2018eaz,Lee:2019tst,Joshi:2019nzi,Marchesano:2019ifh,Font:2019cxq,Lee:2019xtm,Grimm:2019wtx}. 
The {\em refined} swampland distance conjecture was proposed in \cite{Baume:2016psm,Klaewer:2016kiy,Palti:2017elp}. It gives a bound on the lengths of geodesics in the scalar moduli space of a theory of quantum gravity. In \cite{Blumenhagen:2018nts,Blumenhagen:2018hsh} the conjecture was tested by computing the lengths of geodesics in the K\"ahler moduli space $\mathcal{M}_K$ of CY threefolds. This was done by explicitly calculating the (quantum corrected) metric on $\mathcal{M}_K$ and numerically solving the geodesic equation in order to determine the lengths of the geodesics. In particular it was shown that geodesics starting at Landau-Ginzburg points, i.e. limiting points at finite distance in the moduli space, and going all the way to a large volume point satisfy the RSDC. The main examples were CY hypersurfaces with one and two parameters.  

The goal of this work is to extend the discussion of \cite{Blumenhagen:2018nts} to one-parameter CYs that are more exotic. Our focus will be on examples of CYs that have pseudo-hybrid points. Pseudo-hybrid models \cite{Aspinwall:2009qy} are, like Landau-Ginzburg models, at finite distance in the moduli space, but their field theoretic description is not well-understood. They have been linked to singular CFTs, and it has been argued that they do not have a proper limit where gravity decouples. Due to these somewhat mysterious properties, they make good candidates for testing the RSDC. For this purpose we will compute lengths of geodesics from such points to large volume points.

In order to compute the metric on the moduli space we use the gauged linear sigma model (GLSM) \cite{Witten:1993yc} and supersymmetric localization. In \cite{Jockers:2012dk,Gomis:2012wy} it was shown that the sphere partition function \cite{Benini:2012ui,Doroud:2012xw} of the GLSM computes the exact K\"ahler potential on $\mathcal{M}_K$. This method allows us to compute the metric directly on $\mathcal{M}_K$ without taking the detour to the complex structure moduli space of the mirror. Furthermore, by means of supersymmetric localization one can access all regions of $\mathcal{M}_K$ and is not restricted to well-studied large volume settings. Another advantage is that these methods also apply for CYs associated to non-abelian GLSMs. Such CYs cannot be analyzed within the framework of toric geometry. One of our examples will be a one-parameter CY with a pseudo-hybrid point that arises from a GLSM with non-abelian gauge group \cite{Hori:2013gga}.  

The article is organized as follows. In section \ref{sec-distance} we recall the swampland distance conjecture and its refinement and discuss how it is realized for K\"ahler moduli spaces of CYs. Section \ref{sec-glsm} provides a lightning review on GLSMs and the sphere partition function.  In section \ref{sec-models} we discuss our examples and give some details on the properties of pseudo-hybrids from the viewpoint of the GLSM. We then proceed to compute the sphere partition function. The main results of the article are presented in section \ref{sec-results} where we test the RSDC for our examples. We confirm the conjecture holds for these exotic models.\\\\
While this work was in preparation \cite{Joshi:2019nzi} appeared which also computes the metrics on $\mathcal{M}_K$ for the examples we discuss, albeit with different methods. \\\\
{\bf Acknowledgments:} We would like to thank Ralph Blumenhagen, Emanuel Scheidegger, Thorsten Schimannek, Eric Sharpe and Harald Skarke for discussions and comments on the manuscript. JK thanks the University of Chicago for hospitality. The authors were supported by the Austrian Science Fund (FWF): [P30904-N27]. JK was also supported by a faculty grant of the Universities of Chicago and Vienna.
\section{Refined swampland distance conjecture}
\label{sec-distance}
The swampland distance conjecture (SDC) is a statement on the properties of the scalar moduli spaces of a consistent theory of quantum gravity. The claim is that, given a scalar moduli space $\mathcal{M}$ and a point $p_0\in\mathcal{M}$, there exist other points $p\in\mathcal{M}$ that are arbitrarily far away from $p_0$. At these parametrically large distances $\Theta=d(p,p_0)$ an infinitely large tower of light states appears,
\begin{equation}
  \label{massdistance}
  M\sim M_0e^{-\lambda\Theta},
\end{equation}
where $M_0$ and $M$ are the masses at $p_0$ and $p$, respectively. The geodesic distance $\Theta=d(p,p_0)$ is obtained from the metric on $\mathcal{M}$. As $\Theta\rightarrow\infty$ the field theory description breaks down due to  infinitely many massless degrees of freedom. 

In \cite{Klaewer:2016kiy,Blumenhagen:2018nts} a {\em refined} swampland distance conjecture (RSDC) was put forward. It gives constraints on the parameter $\lambda$ in (\ref{massdistance}). Define $\Theta_{\lambda}=\lambda^{-1}$ and denote by $\Theta_0$ the geodesic distance from $p_0$ at which the exponential drop-off sets in. Then the RSDC states that, in Planck units,
\begin{equation}
  \label{refined}
  \Theta_{\lambda}\lesssim\mathcal{O}(1),\qquad\qquad\Theta_0\lesssim\mathcal{O}(1).
\end{equation}
In this work we will test the RSDC in a concrete setting and collect further evidence in its favor. To this end, we choose a framework where also the inspiration for many of these conjectures comes from: type II string compactifications on compact CY threefolds. We consider the parameter space $\mathcal{M}$ to be the K\"ahler moduli space $\mathcal{M}_K$ of the CY. 

In such settings one can test the conjecture by explicitly computing lengths of geodesics in $\mathcal{M}_K$. This is not a trivial task, because the stringy K\"ahler moduli space decomposes into chambers, not all of which contain large volume points. Generic paths on $\mathcal{M}_K$ will cross boundaries between chambers. This amounts to analytic continuation beyond the range of validity of a choice of local coordinates on $\mathcal{M}_K$. Thus, the computation of the geodesic distances between two generic points $p_0$ and $p$ crossing various chambers in $\mathcal{M}_K$ has to be split up into computing geodesics within the individual chambers that are matched along the boundaries. According to the RSDC, one expects that all the distances involved are of order $\mathcal{O}(1)$.

The K\"ahler moduli space of a CY is itself a K\"ahler manifold. To obtain the metric, one computes the K\"ahler potential $K(t,\bar{t})$ on $\mathcal{M}_K$, where $t$ are the K\"ahler moduli. Given $K(t,\bar{t})$, the metric is
\begin{equation}
  g_{t\bar{t}}=\partial_t\partial_{\bar{t}}K(t,\bar{t}).
  \end{equation}
The K\"ahler potential is subject to worldsheet instanton corrections. To compute it, one can either use mirror symmetry or results from supersymmetric localization \cite{Benini:2012ui,Doroud:2012xw,Jockers:2012dk,Gomis:2012wy}. The latter link the K\"ahler potential to the sphere partition function of the GLSM associated to the CY. We will use the GLSM method. The necessary ingredients will be summarized in section \ref{sec-glsm}.     

Once one has the K\"ahler metric, one can compute the distance between two points $p_1$ and $p_2$ in the same chamber via 
\begin{equation}
  \label{distanceint}
  \Theta(p_1,p_2)=\int_{\tau_1(p_1)}^{\tau_2(p_2)}d\tau\,\sqrt{g_{t\bar{t}}(\tau)\pdv{t}{\tau}\pdv{\bar{t}}{\tau}},
\end{equation}
where $\tau$ is an affine parameter. If we take into account quantum corrections to comparatively high orders, it is hard to evaluate this integral analytically.  
Instead, the geodesics can be obtained by numerically solving the geodesic equation 
\begin{equation}
  \label{geodesic}
  \frac{d^2x^{\rho}}{d\tau^2}+\Gamma^{\rho}_{\mu\nu}\frac{dx^{\mu}}{d\tau}\frac{dx^{\nu}}{d\tau}=0.
\end{equation}
In the following, we will restrict ourselves to models with $\mathrm{dim}\mathcal{M}_K=1$. It is convenient to express the complex coordinate $x^{\mu}$ in terms of radial and an angular variable, i.e. $x^{\mu}(\tau)=(r(\tau),\varphi(\tau))$ in (\ref{geodesic}). In these coordinates, the geodesic equation becomes
\begin{align}
    \ddot{\varphi} &= \frac{1}{2} g^{\varphi \varphi} \partial_{\varphi} g_{rr} \dot{r}^2
    - g^{\varphi \varphi} \partial_{r} g_{\varphi \varphi} \dot{r}\dot{\varphi}
    - \frac{1}{2} g^{\varphi \varphi} \partial_{\varphi} g_{\varphi \varphi} \dot{\varphi}^2,\\
    \ddot{r} &= -\frac{1}{2} g^{rr} \partial_{r} g_{rr} \dot{r}^2
    - g^{rr} \partial_{\varphi} g_{rr} \dot{r}\dot{\varphi}
    + \frac{1}{2} g^{rr} \partial_{r} g_{\varphi \varphi} \dot{\varphi}^2.
\end{align}
Given suitable starting values, these equations can be solved numerically in order to obtain the lengths of the geodesics. In the examples we have considered we solved the geodesic equation up to order $\mathcal{O}(r^{50})$. Given these results, we can explicitly evaluate the distances in (\ref{refined}). We start from a point at finite distance in a certain chamber in $\mathcal{M}_K$ and then cross the boundary of the chamber to approach a point at infinite distance. The exponential drop-off becomes significant in the second chamber after traversing a path of length $\approx \Theta_{\lambda}$. Therefore, $\Theta_0$ is the proper distance from the starting point to the chamber boundary, and $\Theta_{\lambda}$ the distance after which the tower of light states appears in the chamber with a limiting point at infinite distance. The full length of the geodesic is then characterized by \cite{Blumenhagen:2018nts}
\begin{equation}
  \label{thetac}
  \Theta_c=\Theta_0+\Theta_{\lambda}=\Theta_0+\frac{1}{\lambda}.
\end{equation}
Conjecturally, this is also of order $\mathcal{O}(1)$. In section \ref{sec-results} we will test the RSDC for such examples. 
\section{GLSM and sphere partition function}
\label{sec-glsm}
To define an $\mathcal{N}=(2,2)$ GLSM we choose a (not necessarily abelian) gauge group $G$. The scalar components $\phi$ of the chiral multiplets are coordinates on a vector space $V$. They transform in the matter representation $\rho_V$. In the general case one has $\rho_V:G\rightarrow GL(V)$. For the CY case the matter representation must satisfy $\rho_V:G\rightarrow SL(V)$. This translates into the condition that the weights $Q_i$ ($i=1,\ldots,\mathrm{dim}V$) of $\rho_V$, i.e. the gauge charges of the matter fields, sum to zero. Furthermore there is a $U(1)$ vector R-symmetry under which the fields $\phi$ transform in the representation $R$. 
GLSMs associated to compact CYs have a non-zero superpotential ${W}(\phi)\in\mathrm{Sym}(V^{\ast})$, which has R-charge $2$.  Let $\mathfrak{g}=\mathrm{Lie}(G)$  and $\mathfrak{t}=\mathrm{Lie}(T)$ be the Lie algebras of $G$ and of a maximal torus $T\subset G$, respectively. The scalar components of the gauge multiplet are denoted by $\sigma\in\mathfrak{g}_{\mathbb{C}}$. We further denote the FI parameters and the ($2\pi$-periodic) theta angles by $\zeta,\theta\in i\mathfrak{t}^{\ast}$, respectively. 
For the calculations using the sphere partition function it is convenient to define $t=2\pi\zeta-i\theta$. This FI-theta parameter is related to the complexified K\"ahler parameter of the CY. We further define a pairing $\langle\cdot,\cdot\rangle:\mathfrak{t}_{\mathbb{C}}^{\ast}\times\mathfrak{t}_{\mathbb{C}}\rightarrow\mathbb{C}$. With $t,Q_i\in i\mathfrak{t}^{\ast}_{\mathbb{C}}$, there are natural pairings $\langle t,\sigma\rangle$ and $\langle Q_i,\sigma\rangle$.

To determine the vacua of the theory one has to find the zeros of the scalar potential
\begin{equation}
  U=\frac{1}{8e^2}|[\sigma,\bar{\sigma}]|^2+\frac{1}{2}\left(|\langle Q_i,\sigma\phi_i\rangle|^2+|\langle Q_i,\bar{\sigma}\phi_i\rangle|^2\right)+\frac{e^2}{2}\left(\mu(\phi)-\zeta\right)^2+|dW(\phi)|^2,
\end{equation}
where $\mu:V\rightarrow i\mathfrak{g}^{\ast}$ is the moment map and $e$ is the gauge coupling. The first term implies $\sigma\in\mathfrak{t}_{\mathbb{C}}$ which we will assume from now on. Among the solutions to $U=0$, there are two special classes. The first one is $\sigma=0$ which implies
\begin{equation}
  \mu(\phi)-\zeta=0,\qquad dW(\phi)=0.
\end{equation}
The first expression are the D-term equations, the second the F-terms. There can be different solutions depending on the values of $\zeta$. These are referred to as {\em phases} of the GLSM. They divide up the parameter space, and hence $\mathcal{M}_K$, into chambers. In this way GLSMs can be used to probe the stringy K\"ahler moduli space.  In the phases some or all of the gauge symmetry is broken. If a continuous subgroup is unbroken, one has a strongly coupled phase, otherwise one refers to the phases as Higgs phases.

Another type of solutions are those where $\phi_i=0$ for all $i$. Then the theory develops a Coulomb branch. Classically, this can only happen if some of the FI parameters are zero, i.e. at phase boundaries. The gauge group is broken to the maximal torus $T$, and the $\sigma$ are unconstrained. The Coulomb branch is lifted by one-loop corrections which generate an effective potential for $\sigma$: \begin{equation}
  \mathcal{W}_{eff}=-\langle {t},\sigma\rangle-\sum_i\langle Q_i,\sigma\rangle\left(\log(\langle Q_i,\sigma\rangle) -1\right)+i\pi\sum_{\alpha>0}\langle\alpha,\sigma\rangle. 
  \end{equation}
The first term is the classical term, $\alpha>0$ are the positive roots of $G$. The Coulomb branch persists at the critical locus of $\mathcal{W}_{eff}$. Via mirror symmetry this locus maps to part of the discriminant in the complex structure moduli of the mirror CY. In models with more than one K\"ahler modulus there are also mixed branches. Since we will not discuss such models in this work, we refrain from giving details. When computing geodesics, one should avoid the singular loci encoded in $\mathcal{W}_{eff}$. 

For the computation of the K\"ahler potential $K(t,\bar{t})$ we use the sphere partition function $Z_{S^2}$. This was first computed in \cite{Benini:2012ui,Doroud:2012xw} via supersymmetric localization. The connection to $K(t,\bar{t})$ was first observed in \cite{Jockers:2012dk}:
\begin{equation}
  Z_{S^2}= e^{-K(t,\bar{t})}.
\end{equation}
The definition of the sphere partition function for the CY case is 
\begin{equation}    \begin{split}
      Z_{S^2}=C\sum_{m\in\mathbb{Z}}\int_{i\mathfrak{t}}d^{\mathrm{rk}G}\sigma \prod_{\alpha>0} (-1)^{\langle \alpha, m \rangle}\left(\frac{\langle\alpha,m\rangle^2}{4}+\langle \alpha,\sigma\rangle^2\right) \\
      \times\prod_{i=1}^{\mathrm{dim}V}\frac{\Gamma\left(\frac{R_i}{2}-i\langle Q_i,\sigma\rangle-\frac{\langle Q_i,m\rangle}{2}\right)}{\Gamma\left(1-\frac{R_i}{2} +i\langle Q_i,\sigma\rangle-\frac{\langle Q_i,m\rangle}{2}\right)} e^{4 \pi i\langle\zeta,\sigma\rangle+i\langle \theta,m\rangle},\label{zd2def}
  \end{split}
\end{equation}
where $C$ is a normalization constant. The integral can be evaluated using the residue theorem. Depending on the phase, entering via $\zeta$ in the exponential, one closes the contours such that one gets a convergent expression. In models with non-abelian $G$, these are multidimensional integrals even if $\mathrm{dim}\mathcal{M}_K=1$. 
\section{Models with $\mathrm{dim}\mathcal{M}_K=1$}
\label{sec-models}
In this section we will discuss examples of CYs with one K\"ahler parameter and their associated GLSMs. The GLSMs have two phases, i.e. the stringy K\"ahler moduli space is divided into two chambers. Along the boundary between two phases there are singular points where the theory has a Coulomb branch. We will mainly focus on geodesics that cross the phase boundary.

The limiting points in the phases and at the phase boundaries of one-parameter models can be characterized by the local exponents ${\bf a}=(a_1,a_2,a_3,a_4)$ ($a_i\in\mathbb{Q}$) associated to each point. These exponents are determined by the Picard-Fuchs operator $\mathcal{L}(z)$ related to the CY. To be precise, the Picard-Fuchs operator is usually associated to the mirror CY where $z$ is a local coordinate in the complex structure moduli space of the mirror. However, it has been shown that also the sphere and the hemisphere partition function of the GLSM satisfy GKZ and Picard-Fuchs equations \cite{Halverson:2013eua,Knapp:2016rec,Gerhardus:2018zwb}. Therefore it makes sense to associate differential operators to phases of GLSMs.

Consider the Picard-Fuchs differential operator $\mathcal{L}(z)$, where $z$ is the local coordinate around the singular point at $z=0$. The exponents are determined by solving
\begin{equation}
  \mathcal{L}(z)z^{\bf a}\sum_{n=0}^{\infty}c_nz^n=0.
\end{equation}
Picard-Fuchs operators of this type have been constructed systematically \cite{MR3822913}. Depending on the structure of ${\bf a}$, the corresponding points in $\mathcal{M}_K$ have been referred to as $F$-, $C$-, $K$-, and $M$-points in \cite{MR3822913,Joshi:2019nzi}.
Their properties are summarized in table \ref{tab-fckm}.
\begin{table}
  \begin{center}
    \begin{tabular}{c|c|c|c}
     {\bf type}&${\bf a}$&{\bf distance on $\mathcal{M}$}&{\bf description}\\
    \hline
    $F$&$(a,b,c,d)$&finite& Landau-Ginzburg, pseudo-hybrid\\
    \hline
    $C$&$(a,b,b,c)$&finite & (mirror of) conifold, pseudo-hybrid\\
    \hline
    $K$&$(a,a,b,b)$&infinite& hybrid\\
    \hline
    $M$&$(a,a,a,a)$&infinite& geometric\\
  \end{tabular}\caption{ $F$-, $C$-, $K$-, and $M$-points of one-parameter models.}\label{tab-fckm}
  \end{center}
  \end{table}
$F$-points are at finite distance in the moduli space. Typical examples are Landau-Ginzburg orbifold theories, but they can also correspond to more exotic phases that have properties similar to Landau-Ginzburg models. 
$C$-points are also at finite distance. These points arise at the singular points at the phase boundaries, but can also appear as limiting points in a phase. The latter case has been studied for instance in \cite{Aspinwall:2009qy,Aspinwall:2015zia}, where models of this type were named pseudo-hybrids. $K$-points are at infinite distance in the moduli space, and it has been argued in \cite{Joshi:2019nzi} that an infinite number of D-branes becomes massless, as expected by the SDC. $K$-points typically correspond to hybrid phases, which are described in terms of Landau-Ginzburg orbifolds fibered over some base manifold. Finally, $M$-points are also at infinite distance. These are geometric points, as one finds them in large volume phases. In the one-parameter case, the RSDC has been shown to hold for geodesics connecting Landau-Ginzburg and $M$-points \cite{Blumenhagen:2018nts}. In this work we will discuss geodesics from $C$-points and other examples of $F$-points to $M$-points. 

Let us give some more details on those $C$-points that are the limiting points in pseudo-hybrid phases. We will characterize pseudo-hybrid phases by their common features. In all the known examples the classical vacuum in the pseudo-hybrid phase has more than one component. The different components exhibit different symmetry breaking patterns.  In contrast to large volume or Landau-Ginzburg phases, a low energy effective description is currently unknown for pseudo-hybrid models. In \cite{Aspinwall:2009qy} it was argued that such models may not have a proper limit where gravity decouples. The associated conformal field theories are singular. Another feature of these phases is that a finite number of branes become massless at the limiting point. Note that in the context of complete intersection CYs also certain $F$-points can exhibit pseudo-hybrid behavior in the sense that the vacuum configurations have several branches associated to different Landau-Ginzburg models.  

In the next three subsections we will introduce the examples we will be working with. We will recall the GLSM description and the discussion of the phases. We will first focus on abelian GLSMs with gauge group $G=U(1)$. There are $14$ models corresponding to $14$ CYs whose associated Picard-Fuchs operators are hypergeometric differential equations. See for instance \cite{Libgober:1993hq,Klemm:1993jj} for partial list and \cite{Doran:2005gu} for a full classification. The complete list has also recently been given in \cite{Joshi:2019nzi}. Three models within this class have $C$-type pseudo-hybrid phases, see \cref{tab:abel-pseudos}. The first example was considered from a GLSM point of view in \cite{Aspinwall:2009qy}. We also discuss the other two models. 
 Furthermore we consider an $F$-type example which is not a standard Landau-Ginzburg model. Finally, we also discuss a $C$-type pseudo-hybrid phase of a non-abelian GLSM, first constructed in \cite{Hori:2013gga}. 
\subsection{$C$-type pseudo-hybrid examples from abelian GLSMs}
\label{sec-pseudo}
In the following we will discuss three abelian GLSMs with a $C$-type pseudo-hybrid phase. In \cref{tab:abel-pseudos} we collect some information on these examples, including the geometry in the large volume ($\zeta\gg0$) phase, the Hodge numbers, and the local exponent ${\bf a}$ in the pseudo-hybrid phase at $\zeta\ll0$. For further information see also \cite{Joshi:2019nzi}.
\begin{table}
    \centering
    \begin{tabular}{c|c|c|c|c}
        {\bf label} & $\zeta\gg0$ & $h^{1,1}$ & $h^{2,1}$ & ${\bf a}$ at $\zeta\ll 0$ \\
        \hline
        C1 & $\mathbb{P}^6[3,2,2]$ & $1$ & $73$ & $\left(\frac{1}{3},\frac{1}{2},\frac{1}{2},\frac{2}{3}\right)$ \\
        C2 & $\mathbb{P}^5[4,2]$ & $1$ & $89$ & $\left(\frac{1}{4},\frac{1}{2},\frac{1}{2},\frac{3}{4}\right)$ \\
        C3 & $\mathbb{P}^5_{1^53}[6,2]$ & $1$ & $129$ & $\left(\frac{1}{6},\frac{1}{2},\frac{1}{2},\frac{5}{6}\right)$ \\
    \end{tabular}
    \caption{$C$-type pseudo-hybrids from $U(1)$ GLSMs.}
    \label{tab:abel-pseudos}
\end{table}
\subsubsection{C1}
\label{sec:ph1}
The GLSM of this model was discussed in \cite{Aspinwall:2009qy} and we repeat the GLSM analysis here. The chiral matter content is 
\begin{equation}
  \begin{array}{c|ccc|c}
    \phi&p_1&p_2&p_3&x_{1},\ldots,x_7\\
    \hline
    U(1)&-3&-2&-2&1\\
    R&2-6q&2-4q&2-4q&2q
    \end{array}
\end{equation}
The gauge invariant superpotential is
\begin{equation}
  W=p_1f_1(x_1,\ldots,x_7)+p_2f_2(x_1,\ldots,x_7)+p_3f_3(x_1,\ldots,x_7),
  \end{equation}
where $(f_1,f_2,f_3)$ are generic homogeneous polynomials of degrees $(3,2,2)$, respectively. The D-term equation is
\begin{equation}
  -3|p_1|^2-2|p_2|^2-2|p_3|^2+\sum_{i=1}^7|x_i|^2=\zeta,
\end{equation}
where $\zeta$ is the FI parameter. The F-terms are
\begin{equation}
  f_i=0,\qquad p_i\frac{\partial f_i}{\partial x_j}=0,\qquad i=1,2,3, \quad j=1,\dots,7.
  \end{equation}
In the $\zeta\gg0$ phase we find a smooth complete intersection of codimension $3$ in $\mathbb{P}^6$ as indicated in \cref{tab:abel-pseudos}:
\begin{equation}
  X_{\zeta\gg0}=\mathbb{P}^6[3,2,2]=\{x_i\in\mathbb{P}^6|f_1=f_2=f_3=0\}.
  \end{equation}
The exotic phase is at $\zeta\ll0$. The D-term excludes the point $p_1=p_2=p_3=0$ and the F-terms imply that $x_1=\ldots=x_7=0$. Thus, the classical vacuum is a weighted $\mathbb{P}^2_{322}$. To see the full picture, we have to turn on classical fluctuations of the $x_i$. First we observe that for generic loci in $\mathbb{P}^2_{322}$ we do not find proper vacua. In this case the gauge symmetry is completely broken. The Landau-Ginzburg model fibered over this base does not lead to a well-defined ground state. The superpotential contains quadratic terms, and hence all the $x_i$ are massive. Thus, we get zero contribution to the central charge, which implies that these vacua are not CY. Furthermore R-symmetry is broken: if all $p_i$ get a VEV and thus must have R-charge zero, there is no consistent way to assign R-charge to the $x_i$ such that the Landau-Ginzburg superpotential has R-charge two. 

There are two special loci on the vacuum manifold which exhibit different behavior. At the point $(p_1,p_2,p_3)=(1,0,0)$ there is an unbroken $\mathbb{Z}_3$, and one recovers a Landau-Ginzburg orbifold with superpotential $W_{LG}=f_1$ in $\mathbb{C}^7/\mathbb{Z}_3$.  There are no quadratic terms in the superpotential, and therefore all $x_i$ are massless. The R-symmetry is preserved: we can assign charge $\frac{2}{3}$ to all $x_i$. This amounts to choosing $q=\frac{1}{3}$ in the table of charges above. The central charge of this Landau-Ginzburg model is $\hat{c}=\frac{7}{3}$, and we do not get a superconformal field theory of a CY. This implies that this Landau-Ginzburg model alone cannot describe the theory at low energies.

A further branch is a curve $\mathcal{C}=(0,p_1,p_2)$, where a $\mathbb{Z}_2$ is preserved and all $x_i$ can be assigned R-charge $1$. This corresponds to $q=\frac{1}{2}$ in the table above. Fibering a $\mathbb{Z}_2$ Landau-Ginzburg orbifold over this curve, the superpotential is quadratic. Naively, one would assume that there are only massive degrees of freedom that do not play a role in the low-energy CFT. Also the R-charge assignment would indicate this. However, the situation is more subtle. Note that we can rewrite the Landau-Ginzburg potential as
\begin{equation}
  W=\sum_{i,j=1}^7x_iA^{ij}(p)x_j,
  \end{equation}
where $A^{ij}$ is a generic $7\times 7$ matrix, linear in $p_{1,2}$. When the rank of $A(p)$ drops, i.e. when $\mathrm{det}A(p)=0$ there will be massless degrees of freedom. This is a situation similar to the models studied in \cite{Hellerman:2006zs,Caldararu:2007tc}.

To summarize, the $\zeta\ll0$-phase has two branches where the gauge symmetry is broken to a $\mathbb{Z}_{2}$ and a $\mathbb{Z}_3$, respectively. The former is of a hybrid-type, the latter is a Landau-Ginzburg orbifold. This is the typical behavior of a pseudo-hybrid model.

Since we are interested in computing geodesics that cross phase boundaries we also have to determine the singularities at the phase boundaries that are encoded in the Coulomb branch of the GLSM. Given the scalar component $\sigma$ of the vector multiplet, the effective potential is
\begin{align}
  \mathcal{W}_{eff}=&-t\sigma-(-3\sigma)\left[\log(-3\sigma)-1\right]-2(-2\sigma)\left[\log(-2\sigma)-1\right]-7\sigma\left[\log\sigma-1\right].
  \end{align}
The critical locus is at
\begin{equation}
  e^{-t}=-\frac{1}{432}\quad\Rightarrow\quad \zeta=\frac{1}{2\pi}\log2^43^3,\quad\theta=\pi\mod 2\pi.
\end{equation}
Now we compute the sphere partition function in the two phases. Inserting into the definition (\ref{zd2def}) gives 
\begin{equation}
\begin{split}
Z_{S^2} = \sum\limits_{m \in \mathbb{Z}}  &\int
\limits^{\infty}_{-\infty} 
\frac{\dd\sigma}{2 \pi} 
e^{(-4 \pi i \zeta \sigma - i \theta m) }\\
&\frac{\Gamma{\left(q- i \sigma - \frac{m}{2}
        \right)}^7}
{\Gamma{\left(1-q+ i \sigma - 
        \frac{m}{2}\right)}^7}
\frac{\Gamma{\left(1-2q+ 2i \sigma + m\right)}^2}
{\Gamma{\left(2q-2 i \sigma +m\right)}^2}
\frac{\Gamma{\left(1-3q+3 i \sigma + 
        \frac{3}{2}m\right)}}
{\Gamma{\left(3q-3 i \sigma + \frac{3}{2}
        m\right)}}.
\end{split}
\end{equation}
Applying the transformation 
\begin{equation}
    \tau = -q + i \sigma \label{eqn:phTrafo}
\end{equation}
 we get 
\begin{equation}
  \begin{split}
    \label{pseudo-z}
Z_{S^2} = \sum\limits_{m \in \mathbb{Z}}  
&\int\limits^{-q+i\infty}_{-q-i\infty} 
\frac{\dd \tau}{2 \pi i} 
\underbrace{e^{-4 \pi  \zeta q}e^{-4 \pi  \zeta \tau - i \theta m }}_{Z_{class}}\\
&\underbrace{\frac{\Gamma{\left(-\tau- \frac{m}{2}\right)}^7}
{\Gamma{\left(1+\tau - \frac{m}{2}\right)}^7}}_{Z_x}
\underbrace{\frac{\Gamma{\left(1+2\tau + m\right)}^2}
{\Gamma{\left(-2\tau +m\right)}^2}}_{Z_{p_2}}
\underbrace{\frac{\Gamma{\left(1+3 \tau + \frac{3}{2}m\right)}}
{\Gamma{\left(-3 \tau + \frac{3}{2}m\right)}}}_{Z_{p_2}}.
\end{split}
\end{equation}
To evaluate the sphere partition function in a given phase we close the integration contour in such a way that the integral converges. One has to take into account the poles of the Gamma functions that are enclosed in the contour. For details of this rather technical discussion we refer to appendix \ref{app-sphere}.

In the $\zeta\gg0$ phase we close the contour to the right so that only the poles coming from $Z_{x}$ contribute. We obtain the following result:
\begin{equation}
  Z_{S^2}^{\zeta\gg0}=-\left(z \bar{z} \right)^{{q}} \operatorname{Res}_{\tau=0} \left(\left(z \bar{z} \right)^{\tau} \pi^4 \frac{\sin \left(2\pi \tau \right)^2 \sin \left(3\pi \tau \right)}{\sin \left(\pi \tau \right)^7} 
    f[\tau,z] \right),
    \label{eqn:phys2lr}
  \end{equation}
where
\begin{equation}
  f[\tau,z] = \left| \sum_{a=0}^{\infty} (-z)^a \frac{\Gamma \left(1+2a+2\tau \right)^2\Gamma \left(1+3a+3\tau \right)}{\Gamma \left(1+a+\tau \right)}\right|^2,
  \end{equation}
and
\begin{equation}
  z = ^{-2\pi \zeta + i \theta}\equiv e^{-t}.
  \label{eqn:ztrafo}
\end{equation}

The evaluation of the sphere partition function in the $\zeta\ll0$ phase is more involved. Closing the contour to the left, two types of poles of $Z_{p_1}$ and $Z_{p_2}$ contribute. Due to coinciding poles we have to be mindful not to over-count. There are two possibilities, namely a pole of $Z_{p_1}$ is a simultaneous pole of $Z_{p_2}$ and vice versa. A careful analysis reveals that whenever this happens the denominator of $Z_x$ cancels such a contribution. Taking this into account, the sphere partition function is a sum of two terms. Again, we refer to the appendix for details. The total result is
\begin{equation}
  Z_{S^2}^{\zeta\ll0}=Z_{S^2,Z_{p_1}}^{\zeta \ll 0}+Z_{S^2,Z_{p_2}}^{\zeta \ll 0},
\end{equation}
where
\begin{equation}
    Z_{S^2,Z_{p_1}}^{\zeta \ll 0} = \sum_{\delta = 0}^{2} \pi^{-4} (-1)^{\delta} \left(z \bar{z}\right)^{{q}- \frac{1+ \delta}{3}} \operatorname{Res}_{\tau=0} \left(\left(z\bar{z}\right)^\tau \frac{\sin \left(\pi \left( \frac{2- \delta}{3}+\tau\right)\right)^7}{\sin \left(3 \pi \tau \right)\sin \left(\pi \left(\frac{1-2 \delta}{3}+2\tau\right)\right)^2} f_1[\tau,z,\delta]\right),
     \label{eqn:phys2lg1}
\end{equation}
with
\begin{equation}
    f_1[\tau,z, \delta] =  \left| \sum_{a=0}^{\infty} (-z)^{-a} \frac{\Gamma\left(\frac{1+\delta}{3}+a-\tau\right)^7}{\Gamma\left(1+3a+\delta-3\tau\right)\Gamma\left(\frac{2+2\delta}{3}+2a -2\tau\right)^2}\right|^2.
\end{equation}
The second contribution is
\begin{equation}
    Z_{S^2,Z_{p_2}}^{r \ll 0} = \left(z \bar{z}\right)^{{q}-\frac{1}{2}} \operatorname{Res}_{\tau=0} \left( \pi^{-4} \left(z \bar{z}\right)^\tau \frac{\sin \left(\pi \left(\frac{1}{2}-\tau\right)\right)^7}{\sin \left(\pi \left(-\frac{1}{2}+3\tau\right)\right)\sin \left(2\tau \pi \right)^2} f_2[\tau,z,0]\right)\label{eqn:phys2lg2}
\end{equation}
with
\begin{equation}
    f_2[\tau,z,\delta] = \left|\sum_{a=0}^{\infty} (-z)^a \frac{\Gamma\left(\frac{1+\delta}{2}+a-\tau\right)^7}{\Gamma\left(
        1+2a+\delta-2\tau\right)^2\Gamma\left(\frac{3+3\delta}{2}+3a -3\tau\right)}\right|^2.
\end{equation}
\subsubsection{C2} 
\label{sec:ph2}
In this model the field content of the $U(1)$ GLSM is 
\begin{equation}
\begin{array}{c|cc|c}
&p_1 & p_2 & x_1\dots x_6 \\ \hline
U(1) & -4 & -2& 1 \\
R & 2-8q & 2-4q & 2q
\end{array}.
\end{equation}
The superpotential is
\begin{equation}
  W=p_1f_4(x_1,\ldots,x_6)+p_2f_2(x_1,\ldots,x_6),
\end{equation}
where $f_4,f_2$ are generic polynomials of degrees $4$ and $2$, respectively.
In the $\zeta\gg0$ phase we recover
\begin{equation}
  X_{\zeta\gg0}=\mathbb{P}^5[4,2]=\{x_i\in\mathbb{P}^5|f_4=f_2=0\}.
  \end{equation}
This complete intersection CY has Hodge numbers $(h^{1,1},h^{2,1})=(1,89)$.

The $\zeta\ll0$ phase is a pseudo-hybrid. The scalar potential is zero for $(p_1,p_2)\in\mathbb{P}^1_{42}$ and $x_i=0$, which does not lead to well-defined vacua for generic $p_1,p_2$. At the point $(p_1,p_2)=(1,0)$ one finds a Landau-Ginzburg orbifold with $W_{LG}=f_4$ in $\mathbb{C}^6/\mathbb{Z}_4$. The R-charge assignment is as above with $q=\frac{1}{4}$. This actually gives a CFT of central charge $\hat{c}=3$, i.e. it is CY. There is another branch with $p_2\neq 0$ where a $\mathbb{Z}_2$ is preserved and the R-charges are given by $q=\frac{1}{2}$. This theory is massive and one is tempted to argue that it does not contribute in the IR. Nevertheless, this branch seems to have some effect, because the limiting point is $C$-point, as is reflected in behavior of $e^{-K(t,\bar{t})}$ computed below. It would be interesting to understand this better. 

The Coulomb branch analysis shows that the singular point is at
\begin{equation}
e^{-t} = \frac{1}{1024} \quad \rightarrow \quad \zeta = \frac{1}{2\pi} \log 2^{10}, \quad \theta = 0 \mod 2 \pi.
\end{equation}
After the transformation (\ref{eqn:phTrafo}) the sphere partition function is
given by 
\begin{equation}
\begin{split}
    Z_{S^2} = \sum_{m\in\mathbb{Z}}\int_{-q - i\infty}^{-q + i \infty} & \frac{\dd \tau}{2\pi i} \underbrace{\frac{\Gamma \left(2 m+4 \tau+1\right)}{\Gamma \left(2 m-4 \tau\right)}}_{Z_{p_1}} 
    \underbrace{\frac{\Gamma \left(m+2 \tau+1\right)}{\Gamma \left(m-2 \tau\right)}}_{Z_{p_2}}\\
    &\underbrace{\frac{\Gamma \left(-\frac{m}{2}-\tau\right)}{\Gamma \left(-\frac{m}{2}+\tau+1\right)}}_{Z_x}
    e^{-4 \pi \zeta q}e^{-i \theta m-4 \pi \zeta \tau}.
\end{split}
\label{eqn:sppf_ph2}.
\end{equation}
The evaluation of the partition function is similar to the 
 C1-case. Therefore we will simply state the results
in the two phases.

For $ \zeta \gg 0$ on the poles of $Z_x$ in (\ref{eqn:sppf_ph2})
contribute and we obtain
\begin{align}
Z_{S^2}^{r \gg 0} = -(z\bar{z})^q\pi^4 \operatorname{Res}_{\tau = 0} \left( \frac{\sin 4\pi \tau \sin 2 \pi \tau}{(\sin\pi \tau)^6} (z\bar{z})^{\tau} \left| f[z,\tau]\right|^2\right),
\label{eqn:sppf_ph2_final}
\end{align}
with
\begin{align}
f[z,\tau] = \sum_{n=0}^{\infty} z^n \frac{\Gamma \left(1+4n+4\tau\right)\Gamma\left(1+2n+2\tau\right)}{\Gamma\left(1+n+\tau \right)^6}.
\end{align}
The steps to obtain the result in the phase $\zeta \ll 0$ 
are the same as for the previous example. Again, we find two 
contributions from $Z_{p_1}$ and $Z_{p_2}$:
\begin{equation}
    Z_{S^2}^{\zeta \ll 0} =  Z_{S^2, Z_{p_1}}^{\zeta \ll 0} + Z_{S^2, Z_{p_2}}^{\zeta \ll 0}.
\end{equation}
However, in contrast to the previous example we now have to take into account a possible over-counting of poles. Details on this issue are given in \cref{sec:ap_pph2}. From the poles of $Z_{p_2}$ we get the following contributions:
\begin{align}
Z^{\zeta \ll 0}_{S^2,Z_{p_2}} =   (z\bar{z})^{q- \frac{1}{2}} \operatorname{Res}_{\tau =0 } \left(- \pi^{-4} \frac{\left(\sin\left(\left(\frac{1}{2} + \tau \right) \pi\right)\right)^6}{\sin 4 \pi \tau \sin 2 \pi \tau}(z \bar{z})^{\tau}  \left|\tilde{f}_1[z,\tau]\right|^2\right),
\end{align}
with
\begin{align}
\tilde{f}_1 [z,\tau] = \sum_{a=0}^{\infty} z^{-a}\frac{\Gamma \left(a + \frac{1}{2} -\tau\right)^6}{\Gamma \left(4a +2-4 \tau\right)\Gamma \left(2a +1 -2 \tau\right)}.
\end{align}
From the poles of $Z_{p_1}$ we get
\begin{align}
Z_{S^2,Z_{p_1}}^{\zeta \ll 0} = \sum_{\delta =0,2}  (z \bar{z})^{q- \frac{\delta+1}{4}} (-1)^{\delta} \operatorname{Res}_{\tau = 0} \left( \pi^{-4} \frac{\sin \left(\pi\left(\frac{3- \delta}{4}+\tau\right)\right)^6}{\sin 4\pi \tau \sin \left(\pi\left(\frac{1-\delta}{2}+2\tau\right)\right)} (z \bar{z})^{\tau} \left| \tilde{f}_2[z,\tau,\delta]\right|^2\right),
\end{align}
with
\begin{align}
\tilde{f}_2[z,\tau,\delta] = \sum_{a=0}^{\infty} z^{-a} \frac{\Gamma \left(a + \frac{\delta +1}{4} -\tau\right)^6}{\Gamma \left(1+4a+\delta -4\tau \right)\Gamma \left(2a + \frac{\delta +1}{2}-2\tau\right)}.
\end{align}
\subsubsection{C3}
\label{sec:ph3}
As the final example of this class we consider the $U(1)$ GLSM with the following field content
\begin{align}
\begin{array}{c|cc|cc}
& p_1 & p_2 & x_1,\ldots,x_5 & x_6 \\ 
\hline
U(1) & -6 &-2 & 1 & 3 \\
R & 2-12q & 2-4q & 2q &6q
\end{array}
\end{align}
The superpotential is
\begin{equation}
  W=p_1f_6(x_1,\ldots,x_6)+p_2f_2(x_1,\ldots,x_5),
\end{equation}
with generic polynomials $f_6,f_2$ of degrees $6$ and $2$, respectively. The $\zeta\gg0$ phase is again a complete intersection
\begin{equation}
  X_{\zeta\gg0}=\mathbb{P}^5_{1^53}[6,2]=\{x_i\in\mathbb{P}^5_{1^53}|f_6=f_2=0\},
\end{equation}
with Hodge numbers $(h^{1,1},h^{2,1})=(1,129)$. The other phase is a pseudo-hybrid. The classical potential is zero for $(p_1,p_2)\in\mathbb{P}^1_{62}$ and $x_i=0$. At $(p_1,p_2)=(1,0)$ one finds a Landau-Ginzburg orbifold in $\mathbb{C}^7/\mathbb{Z}_6$ with potential $W_{LG}=f_6$, massive $x_6$, and R-charge assignment $q=\frac{1}{6}$. Curiously, the central charge of the CFT is $\hat{c}=\frac{10}{3}$, which exceeds the value for the CY case. There is another branch with $p_2\neq 0$ with an unbroken $\mathbb{Z}_2$ and R-charges given by $q=\frac{1}{2}$. The locus of the singularity is 
\begin{equation}
e^{-t} = \frac{1}{6912} \quad \rightarrow \quad \zeta = \frac{1}{2\pi} \log 3^32^8, \quad \theta = 0 \mod 2 \pi.
\end{equation}
After the transformation (\ref{eqn:phTrafo}) the sphere partition function 
reads
\begin{equation}
    \begin{split}
        Z_{S^2} =  \sum_{m\in\mathbb{Z}}\int & \frac{\dd \tau}{2\pi i} \underbrace{\frac{\Gamma \left(3 m+6 \tau+1\right)}{\Gamma \left(3 m-6 \tau\right)}}_{Z_{p_1}} \underbrace{\frac{\Gamma \left(m+2 \tau+1\right)}{\Gamma \left(m-2 \tau\right)}}_{Z_{p_2}} \\
        &\underbrace{\frac{\Gamma \left(-\frac{m}{2}-\tau\right)^5}{\Gamma \left(-\frac{m}{2}+\tau+1\right)^5}}_{Z_{x}}
        \underbrace{\frac{\Gamma \left(-\frac{3}{2}m-3 \tau\right)}{\Gamma \left(-\frac{3 m}{2}+3 \tau+1\right)}}_{Z_{\tilde{x}}}
        e^{-4 \pi  \zeta \left(q+\tau\right)-i \theta  m}. \label{eqn:sppf_ph3}
    \end{split}
\end{equation}
The evaluation in the different phases can be done by the same methods
as in the previous models and therefore we give only the final results in
the two phases. The only subtlety is that now also in the $\zeta \gg 0$ phase
two different poles contribute and we must take into account a potential over-counting. This
is similar to the situation for the model in \cref{sec:ph2} and therefore can 
be treated in the same way as outlined in \cref{sec:ap_pph2}.

In the $\zeta \gg 0 $ phase we find that the only contribution is given by
coinciding poles of $Z_{x}$ and $Z_{\tilde{x}}$:
\begin{align}
Z_{S^2}^{\zeta \gg 0} =-(z \bar{z})^q \operatorname{Res}_{\tau=0} \left( \pi^4(z\bar{z})^{\tau} \frac{\sin 6\pi \tau \sin 2 \pi \tau}{\left(\sin \pi \tau\right)^5\sin3\pi \tau} \left| f_{1} [z,\tau]\right|^2 \right) ,
\end{align}
with
\begin{align}
f_1[z,\tau] = \sum_{a=0}^{\infty} z^a \frac{ \Gamma\left(1+6a +6\tau \right)\Gamma \left(1+2a+2\tau\right) }{\Gamma \left(1+a+\tau\right)^5\Gamma \left(1+3a+3\tau\right)}.
\end{align}  
For $ \zeta \ll 0$ we again find two contributions:
\begin{align}
Z^{\zeta \ll 0}_{S^2,Z_{p_2}} =  (z \bar{z})^{q- \frac{1}{2}} \operatorname{Res}_{\tau =0} \left( \pi^{-4} (z\bar{z})^{\tau}\frac{\left(\sin \left(\pi \left( \frac{1}{2}+\tau\right)\right)\right)^5\sin \left(\pi\left(- \frac{1}{2} +3 \tau\right)\right)}{\sin 6 \pi \tau \sin 2 \pi \tau} \left|\tilde{f}_1 [z,\tau]\right|^2\right),
\end{align}
with
\begin{align}
\tilde{f}_1[z,\tau] = \sum_{a=0}^{\infty} z^{-a} \frac{\Gamma\left(\frac{1}{2}+a-\tau\right)^5\Gamma \left(\frac{3}{2}+3a-3\tau\right) }{\Gamma\left(3+ 6a  -6\tau \right)\Gamma\left(1+2a  -2 \tau\right)}.
\end{align}

The second contribution is given by:
\begin{multline}
Z_{S^2,Z_{p_1}}^{\zeta \ll 0} = \sum_{\delta \in \{0,1,3,4 \}} (z \bar{z})^{q - \frac{1+\delta}{6}} (-1)^{\delta} \\ 
\operatorname{Res}_{\tau = 0} \left( \pi^{-4} (z \bar{z})^{\tau} \frac{\left(\sin \left(\pi \left(\frac{5-\delta}{6}+\tau\right)\right)\right)^5\sin \left(\pi \left( \frac{1-\delta}{2}+3\tau\right)\right)}{\sin 6 \pi \tau\sin \left(\pi\left(\frac{1+\delta}{3}-2\tau\right)\right)} \right.\\ 
\left. \left| \tilde{f}_2[z,\tau,\delta]| \right|^2\right),
\end{multline}
and
\begin{align}
\tilde{f}_2[z,\tau,\delta] = \sum_{a=0}^{\infty} z^{-a} \frac{\Gamma\left(\frac{\delta+1}{6}+a-\tau\right) ^5\Gamma\left(\frac{1+\delta}{2}+3a -3 \tau\right)}{\Gamma \left(\delta+1 + 6 a -6\tau\right)\Gamma \left(\frac{1+\delta}{3}+2a -2 \tau\right)}.
\end{align}
Closer inspection reveals that the poles cancel except for $\delta \in \{0,4\}$.
\subsection{$F$-type examples from abelian GLSMs}
\label{sec-ftype}
Among the $14$ hypergeometric examples there are three with limiting points of type $F$ that do not correspond to Landau-Ginzburg points. Earlier discussions of these examples can be found in \cite{Klemm:1993jj,Hosono:1994ax,Klemm:2004km,MR3409874,Joshi:2019nzi}. We collect the relevant information about these models in  \cref{tab:abel-ftype}.
\begin{table}
    \centering
    \begin{tabular}{c|c|c|c|c|c}
        {\bf label} & $\zeta\gg0$: $\mathbb{P}^5_{w_i}[d_1,d_2]$ & $h^{1,1}$ & $h^{2,1}$ & ${\bf a}$ at $\zeta\ll 0$ & $\mu$ (sing. at $e^{-t}={\mu}^{-1}$) \\
        \hline
        F1 & $\mathbb{P}^5_{1^52}[4,3]$ & $1$ & $79$ & $\left(\frac{1}{4},\frac{1}{3},\frac{2}{3},\frac{3}{4}\right)$ & $-2^63^3$\\
        F2 & $\mathbb{P}^5_{1^32^23}[6,4]$ & $1$ & $79$ & $\left(\frac{1}{6},\frac{1}{4},\frac{3}{4},\frac{5}{6}\right)$ & $2^{10}3^3$\\
        F3 & $\mathbb{P}^5_{1^44,6}[12,2]$ & $1$ & $243$ & $\left(\frac{1}{12},\frac{5}{12},\frac{7}{12},\frac{11}{12}\right)$ & $2^83^3$\\
    \end{tabular}
    \caption{$F$-type pseudo-hybrids from $U(1)$ GLSMs.}
    \label{tab:abel-ftype}
\end{table}
We describe them in terms of $U(1)$ GLSMs with field content
\begin{equation}
\begin{array}{c|cc|c}
&p_1 & p_2 & x_i \\ \hline
U(1) & -d_1 & -d_2& w_i \\
R & 2-2d_1q & 2-d_2q & 2w_iq
\end{array},
\end{equation}
where the $w_i$  and $d_{1},d_2$ can be read off from  \cref{tab:abel-ftype}. The superpotential has the form
\begin{equation}
  W=p_{1}f_{d_1}(x_i)+p_2f_{d_2}(x_i),
\end{equation}
with $f_{d_1},f_{d_2}$ homogeneous polynomials of degrees $(d_1,d_2)$. The $\zeta\gg0$ phases are the complete intersections indicated in  \cref{tab:abel-ftype}. The vacuum manifold in the $\zeta\ll0$-phase is $(p_1,p_2)\in\mathbb{P}^1_{d_1d_2}$. At the point $(p_1,p_2)=(1,0)$ there is an unbroken $\mathbb{Z}_{d_1}$ and one finds a corresponding Landau-Ginzburg orbifold at this point with a R-charge assignment $q=\frac{1}{d_1}$ for the fields. Analogously, one finds a $\mathbb{Z}_{d_2}$ Landau-Ginzburg orbifold at $(p_1,p_2)=(0,1)$. The singular point in the moduli space and the corresponding values for $\zeta,\theta$ can be read off from the last column in  \cref{tab:abel-ftype}.

Since these models are very similar to the $C$-type models, we only compute the sphere partition function for one of them, namely F1. The definition of the sphere partition function for this case is
\begin{equation}
\begin{split}
  Z_{S^2} = \frac{1}{2 \pi i} \sum_{m\in\mathbb{Z}}\int_{-q - i\infty}^{-q + i \infty} & \dd \tau \underbrace{\frac{\Gamma\left(-\tau-\frac{m}{2}\right)^5}{\Gamma\left(1+\tau-\frac{m}{2}\right)^5}}_{Z_{x_{1\dots5}}}\underbrace{\frac{\Gamma\left(-2\tau-m\right)}{\Gamma\left(1+2\tau-m\right)}}_{Z_{x_6}}\\
    &\underbrace{\frac{\Gamma\left(1+4\tau+2m\right)}{\Gamma\left(-4\tau+2m\right)}}_{Z_{p_1}}\underbrace{\frac{\Gamma\left(1+3\tau+\frac{3m}{2}\right)}{\Gamma\left(-3\tau+\frac{3m}{2}\right)}}_{Z_{p_2}}
    e^{-4 \pi \zeta q}e^{-i \theta m-4 \pi \zeta \tau},
\end{split}
\end{equation}
where we have defined $\tau$ via (\ref{eqn:phTrafo}) as usual. The evaluation in the $\zeta\gg0$-phase is completely analogous to the $C$-type examples. The result is
\begin{equation}
Z_{S^2}^{\zeta \gg 0} =- (z \bar{z})^{q } 
\operatorname{Res}_{\tau =0} \left((z\bar{z})^{\tau} \pi^4  \frac{\sin\left(4 \pi \tau \right) \sin \left( 3 \pi \tau\right)}{\sin\left(\pi \tau \right)^5 \sin\left(2 \pi \tau\right)} \left| f[z,\tau,0]\right|^2\right),
\end{equation}
with
\begin{equation}
f[z,\tau] = \sum_{a=0}^{\infty} (-1)^a z^a \frac{\Gamma \left(4 a+4 \tau+1\right)\Gamma \left(3 a+3 \tau+1\right)}{\Gamma \left(a+\tau+1 \right)^5\Gamma\left(2 a+2 \tau+1\right)}.
\end{equation}
In the $\zeta\ll0$-phase only first order poles contribute. This is a typical behavior for $F$-type models. A study of 
the location of the poles reveals that certain poles of $Z_{p_2}$ are a subset
of the poles of $Z_{p_1}$. Therefore we will first take into account  all poles of $Z_{p_1}$ and then consider the remaining poles of $Z_{p_2}$. We find for the 
$Z_{p_1}$ contribution:
\begin{equation}
\begin{split}
Z_{S^2,Z_{p_1}}^{\zeta \ll 0} = \sum_{\delta=0}^{3} &(-1)^{\delta} \left(z\bar{z}\right)^{q- \frac{\delta+1}{4}} \pi^{-4} \\
&\operatorname{Res}_{\tau =0} \left( \left(z \bar{z}\right)^{\tau}   \frac{\sin \left(\pi \left(\frac{3-\delta}{4}+\tau \right)\right)^5\sin \left(\pi \left(\frac{1-\delta}{2} +2 \tau \right)\right)}{ \sin \left(4 \pi  \tau\right)\sin \left(\pi  \left(\frac{1-3 \delta }{4}+3 \tau\right)\right)}\left|f_1[z,\tau,\delta]\right|^2\right).
\end{split}
\end{equation}
We see that only $ \delta \in \{0,2\}$ gives a non-zero contribution. The 
contributions for $\delta \in \{1,3\}  $ get canceled. This is anticipated as we expect only first order poles in this phase.  Since we only have first order poles, the residue can be evaluated explicitly. We find
\begin{equation}
Z_{S^2,Z_{p_1}}^{\zeta \ll 0} = \frac{\left(z \bar{z}\right)^{q-\frac{3}{4}} \left(\left|f_1[z,0,0]\right|^2 \sqrt{z \bar{z}} -\left|f_1[z,0,2]\right|^2\right)}{16 \pi ^5},
\end{equation}
with
\begin{align}
f_1[z,0, 0] &= \sum_{a=0}^{\infty} (-z)^{-a} \frac{\Gamma \left(
    a+\frac{1}{4}\right)^5\Gamma\left(  2
    a+\frac{1}{2}\right)}{\Gamma \left(4 a +1\right)\Gamma \left( 3a+\frac{3}{4}\right)}. \\
f_1[z,0, 2] &= \sum_{a=0}^{\infty} (-z)^{-a} \frac{\Gamma \left(
    a+\frac{3}{4}\right)^5\Gamma\left(  2
    a+\frac{3}{2}\right)}{\Gamma \left(4 a+3\right)\Gamma \left( 3a+\frac{9}{4}\right)}. 
\end{align} 
The contribution from $Z_{P_2}$ results in:
\begin{equation}
\begin{split}
Z_{S^2,Z_{p_2}}^{\zeta \ll 0} = -\sum_{\delta=0}^{1}&(-1)^\delta \left(z \bar{z} \right)^{q-\frac{1+\delta}{3}} \pi^{-4} \\
&\operatorname{Res}_{\tau =0} \left( (z\bar{z}^{\tau} \frac{\sin \left(\pi  \left(\frac{1+\delta}{3} - \tau\right)\right)^5\sin \left(\pi\left(\frac{1-2  \delta}{3} +2  \tau\right)\right)}{\sin \left( \pi  \left(\frac{1+4 \delta}{3} -4 \tau\right)\right) \sin \left(3 \pi  \tau\right)}\left|f_2[z,\tau, \delta]\right|^2\right).
\end{split}
\end{equation}
Again we only have first order poles and the result for the residue is
\begin{equation}
    Z_{S^2,Z_{p_2}}^{\zeta \ll 0} =-\frac{3 \sqrt{3} \left(z \bar{z}\right)^{q-\frac{2}{3}} \left(\left|f_2[z,0,0]\right|^2 \sqrt[3]{z \bar{z}}-\left|f_2[z,0,1]\right|^2\right)}{32 \pi ^5},
\end{equation}
with
\begin{align}
f_2 [z,0,0] &= \sum_{a=0}^{\infty} (-z)^{-a} \frac{\Gamma\left(
    a+\frac{1}{3}\right)^5\Gamma
    \left(2 a+\frac{2}{3}\right) }{\Gamma  \left(4a+\frac{4}{3} \right)\Gamma \left(3 a+1\right)},\\
f_2 [z,0,1] &= \sum_{a=0}^{\infty} (-z)^{-a} \frac{\Gamma\left(
    a+\frac{2}{3}\right)^5\Gamma
    \left(2 a+\frac{4}{3}\right) }{\Gamma  \left(4a+\frac{8}{3} \right)\Gamma \left(3 a+2\right)}.
\end{align}
We have excluded the case $\delta = 2 $ as this would lead to a
double counting of poles. Nevertheless this contribution would have been
canceled anyway.

\subsection{Non-abelian example with a pseudo-hybrid phases}
\label{sec-nonab}
So far, we have only discussed models that can be realized in terms of toric geometry and thus by abelian GLSMs. All the methods that we have used also apply to non-abelian GLSMs. Indeed, $C$-type pseudo-hybrid phases also appear for non-abelian models. The first one-parameter example with a pseudo-hybrid phase was found in \cite{Hori:2013gga}. Further examples with $\mathrm{dim}\mathcal{M}_K=1$ have been identified in \cite{Caldararu:2017usq}.

Due to conceptual and technical challenges, non-abelian models are difficult to come by. One problem is that one of the phases is typically strongly coupled, i.e. in the low-energy effective theory a {\em continuous} subgroup of the GLSM gauge group is unbroken. While it is understood how to compute the sphere partition function in such cases, the result will not be absolutely convergent, and one needs to apply Borel summation to get a convergent expression. In practice, this is doable only for the very lowest orders in the expansion. Another problem is related to the fact that all known one-parameter non-abelian GLSMs have more than one singular point at the phase boundary. This also means that the associated Picard-Fuchs differential operators are no longer hypergeometric. In the region between the singular points it is not clear whether one should choose the metric for the $\zeta\gg0$-phase or for the $\zeta\ll0$-phase. Neither will have good convergence properties there. A further complication is related to numerics. The expressions for the sphere partition function are more complicated than in the abelian case. This also complicates the numerical calculation of the geodesics. One aspect is that the calculations work much more efficiently for a suitable choice of coordinate on $\mathcal{M}_K$. As has been demonstrated in \cite{Blumenhagen:2018nts} for the abelian examples, it is best to choose a coordinate $\psi$ in such a way that the (only) singular point at the phase boundary is at $\psi=1$. Having more than one singular point, there is no obvious choice for a ``good'' coordinate.

In view of these difficulties, we have not been able to give a complete discussion of geodesics crossing phase boundaries. Nevertheless, we have been able to make some explicit computations by restricting ourselves to geodesics within a weakly-coupled pseudo-hybrid phase. The example we consider is a one-parameter non-abelian GLSM with gauge group $G=U(2)$ that has been discussed in \cite{Hori:2013gga}. The matter content is 
\begin{equation}
\begin{array}{c|cc|cc}
\phi &p^1,\ldots,p^5 & p^6,p^7& x_1,x_2& x_3,\ldots,x_5 \\ \hline
U(2) & \mathrm{det}^{-1} &\mathrm{det}^{-2}& \mathrm{det}\otimes{\Box}&{\Box} \\
R &  4q & 8q & 1-6q & 1-2q
\end{array},
\end{equation}
where $\Box$ refers to the fundamental representation and $\det$ is the determinantal representation of $U(2)$. 
The superpotential is
\begin{equation}
  W=\sum_{i,j=1}^{5}A^{ij}(p)[x_ix_j],
\end{equation}
where $[x_ix_j]=\varepsilon_{ab}x_i^ax_j^b$ ($a,b=1,2$). By gauge invariance, the antisymmetric $5\times 5$ matrix $A(p)$ has the following structure. The first $2\times 2$ block transforms according to $\det^{-3}$, i.e. it is cubic in $p^1,\ldots,p^5$ and bilinear in $(p^{1,\ldots,5},p^{6,7})$. The lower $3\times 3$ block transforms in the $\mathrm{det}^{-1}$ representation, which means that the entries are linear in $p^{1,\ldots,5}$. The off-diagonal blocks are in $\mathrm{det}^{-2}$ representation, i.e. the entries are quadratic in $p^{1,\ldots,5}$ and linear in $p^{6,7}$. 

This model has an $M$-point in the $\zeta\ll0$-phase which is a smooth Pfaffian CY in weighted $\mathbb{P}^7$ \cite{Kanazawa:2012xya} characterized by the condition $\mathrm{rk}A(p)=2$. It is a strongly coupled phase where an $SU(2)$ subgroup of $G$ remains unbroken. The $\zeta\gg0$-phase is a pseudo-hybrid phase of type $C$ that is at finite distance in the moduli space.

There are two singular points at
\begin{equation}
  \label{nonab-sing}
  e^{-t_{\pm}}= (540 \pm 312 \sqrt{3}).
\end{equation}
Note that the points are not at the same theta angle: $\theta_+=0$ and $\theta_-=\pi$ ($\mathrm{mod}\:2\pi$).

The sphere partition function for this model has been computed in \cite{Hori:2013gga}. In our limited discussion we will compute the lengths $\Theta_0$ of geodesics starting at the pseudo-hybrid point and ending at the $\zeta$-value of the nearest singular point. For this purpose we only need the result of the sphere partition function in the $\zeta\gg0$ phase. The sphere partition function can be written as
\begin{align}
  \label{znonab}
Z_{S^2}  =-\frac{1}{8\pi}\int_{\gamma+i\mathbb{R}^2} \left(Z_1\right)^5\left(Z_2\right)^2\left(Z_3\right)^2
\left(Z_4\right)^2\left(Z_5\right)^3
\left(Z_6\right)^3 Z_G Z_{\text{classical}}\:d\tau_1 \wedge d\tau_2,
\end{align}
where we have defined $\tau_i=-q-i\sigma_i$ ($i=1,2$) and $\gamma=-q(1,1)^T$. The contributions to the integrand are
\begin{align}
Z_1&=Z_{p^{1,\dots,5}} = \frac{\Gamma \left(-\tau_1 - \tau_2
+ \frac{1}{2} (m_1+m_2 )\right)}{\Gamma (1+ \tau_1 + \tau_2
+ \frac{1}{2} (m_1+m_2 ))},\\
Z_2&=Z_{p^{6,7}} = \frac{\Gamma \left(-2\tau_1 - 2 \tau_2
+  (m_1+m_2 )\right)}{\Gamma (1+2\tau_1 + 2 \tau_2
+  (m_1+m_2 ))}, \\
Z_3 Z_4 &=Z^1_{x_{1,2}}Z^2_{x_{1,2}} = \frac{\Gamma \left(\frac{1}{2}+2 \tau_1 +  \tau_2
- \frac{1}{2} (2m_1+m_2 )\right)}{\Gamma (\frac{1}{2}-2 \tau_1 - \tau_2
-  \frac{1}{2}(2m_1+m_2 ))}\frac{\Gamma \left(\frac{1}{2}+ \tau_1 +2 \tau_2
- \frac{1}{2} (m_1+2m_2 )\right)}{\Gamma (\frac{1}{2}- \tau_1 -2 \tau_2
-  \frac{1}{2}(m_1+2m_2 ))}, \\
Z_5 Z_6&=Z^1_{x_{3,4}}Z^2_{x_{3,4}} = \frac{\Gamma \left(\frac{1}{2}+\tau_1 - \frac{1}{2} m_1\right)}{\Gamma \left(\frac{1}{2}- \tau_1 - \frac{1}{2} m_1\right)} \frac{\Gamma \left(\frac{1}{2}+\tau_2 - \frac{1}{2} m_2\right)}{\Gamma \left(\frac{1}{2}-\tau_2 - \frac{1}{2} m_2\right)},\\
Z_G &= (-1)^{m_1-m_2} \left(\frac{1}{4} (m_1-m_2)^2 - (\tau_1 -\tau_2)^2\right),\\
Z_{\text{Classical}} &= e^{8 \pi \zeta  q } e^{4 \pi r (\tau_1 + \tau_2) - i \theta (m_1+ m_2)}.
\end{align}
In the $\zeta\ll0$-phase the sphere partition function can be written as
\begin{align}
Z_{S^2 }^{ \zeta \gg 0} = Z_{S^2,1 }^{ \zeta \gg 0} +2Z_{S^2,2 }^{ \zeta \gg 0}.
\end{align}
The first term is
\begin{align}
    Z_{S^2,1}^{\zeta \gg 0} =-\frac{9 \sqrt{3} \left(z \bar{z}\right)^{\frac{1}{3}-2 q} \left(\sqrt[3]{z \bar{z}}\partial_{x_2}^2\left|f_1[z,x_1,x_2,1]\right|^2|_{(0,0)} -\partial_{x_2}^2\left|f_1[z,x_1,x_2,0]\right|^2|_{(0,0)} \right)}{256 \pi ^7},
\end{align}
with
\begin{align}
\begin{split}
f_1[z,x_1,&x_2,\delta] =\sum_{l=0}^{\infty} (-z)^{l} \sum^{3l+\delta}_{b=0} (-1)^b  \left(-2 b+\delta +3 l-x_1+x_2\right)\\
& \frac{\Gamma\left( l+\frac{\delta+1}{3} - x_1-
   x_2\right)^5\Gamma  \left(2 l+\frac{2}{3}(\delta +1)-2 x_1-2
   x_2\right)^2\Gamma\left( -b+l+\frac{\delta }{3}+x_2+\frac{1}{3}\right)^3}{\Gamma \left(-b+3 l+\delta -2 x_1-x_2+1\right)^2\Gamma \left(b-x_1-2 x_2+1\right)^2\Gamma \left(-b+2 l+\frac{2+2 \delta }{3}-x_1\right)^3}.
\end{split}
\end{align}
The second contribution is
\begin{align}
    Z_{S^2,2}^{\zeta\gg 0} =\frac{1}{8} \left(z \bar{z}\right)^{\frac{1}{2}-2 q} \left(3 \left|f_2[z,0,0]\right|^2 \log \left(z \bar{z}\right)-4 \partial_{x_2}\left|f_2[z,x_1,x_2]\right|^2|_{(0,0)}+\partial_{x_1} \left|f_2[z,x_1,x_2]\right|^2|_{(0,0)}\right),
\end{align}
where
\begin{align}
\begin{split}
f_2[z,x_1&,x_2] = \sum_{a=0}^\infty z^{a}\sum_{b=0}^{2a} \left(-2 b+a -x_1+x_2-\frac{1}{2}\right) \\
&\frac{\Gamma\left(a-x_1-x_2+\frac{1}{2}\right)^5\Gamma\left(2 a-2 x_1-2 x_2+1 \right)^2}{\Gamma \left(-b+2 a-2 x_1-x_2+1\right)^2\Gamma \left(b+a-x_1-2 x_2+\frac{3}{2}\right)^2\Gamma \left(-b+a-x_1+\frac{1}{2}\right)^3\Gamma \left(b-x_2+1\right)^3}.
\end{split}
\end{align}
To obtain this result, one has to compute a multi-dimensional residue, which is considerably harder that the one-dimensional case that we had to deal with for the abelian GLSMs. Prescriptions for evaluating such integrals can be found for instance in \cite{russians,Friot:2011ic,Gerhardus:2015sla,Larsen:2017aqb}. We outline some of the steps in appendix \ref{app-nonab}. The leading behavior of the sphere partition function is \cite{Hori:2013gga}
\begin{equation}
 Z_{S^2}^{\zeta\gg0}=\frac{\Gamma \left(\frac{1}{3}\right)^{10} \left(z \bar{z}\right)^{-2 q+\frac{1}{3}}}{2\sqrt{3} \pi  \Gamma \left(\frac{2}{3}\right)^8}- \left(z \bar{z}\right)^{-2 q+\frac{1}{2}} \left(-3 \log \left(z \bar{z}\right)+36+8 \log (4)\right)+ \ldots
\end{equation}
This is the expected behavior for a pseudo-hybrid phase.
\section{Testing the refined swampland distance conjecture}
\label{sec-results}
In this section we present our results for the lengths of geodesics in the models discussed in section \ref{sec-models}. 
\subsection{$C$-type pseudo-hybrid examples from abelian GLSMs}
    \subsubsection{C1}
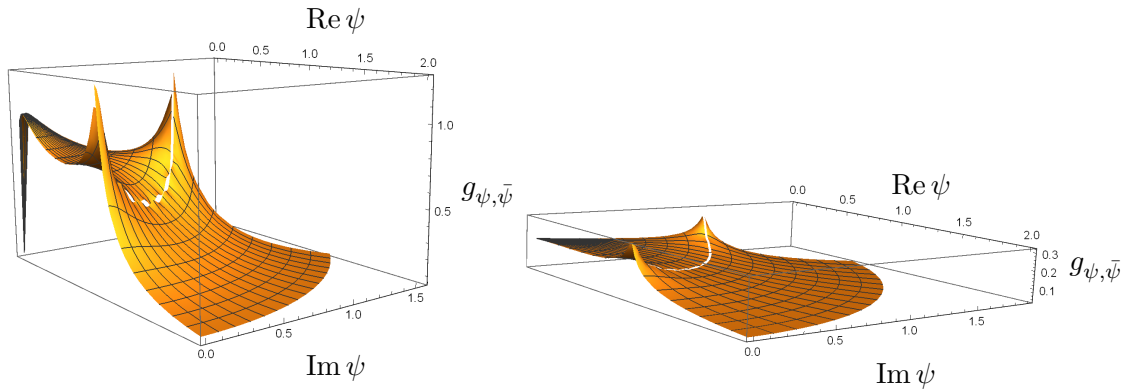
\begin{figure}
        \centering
            \begin{tikzpicture}
            \node [inner sep=0pt,above right] 
            {\includegraphics[width=0.4\textwidth]{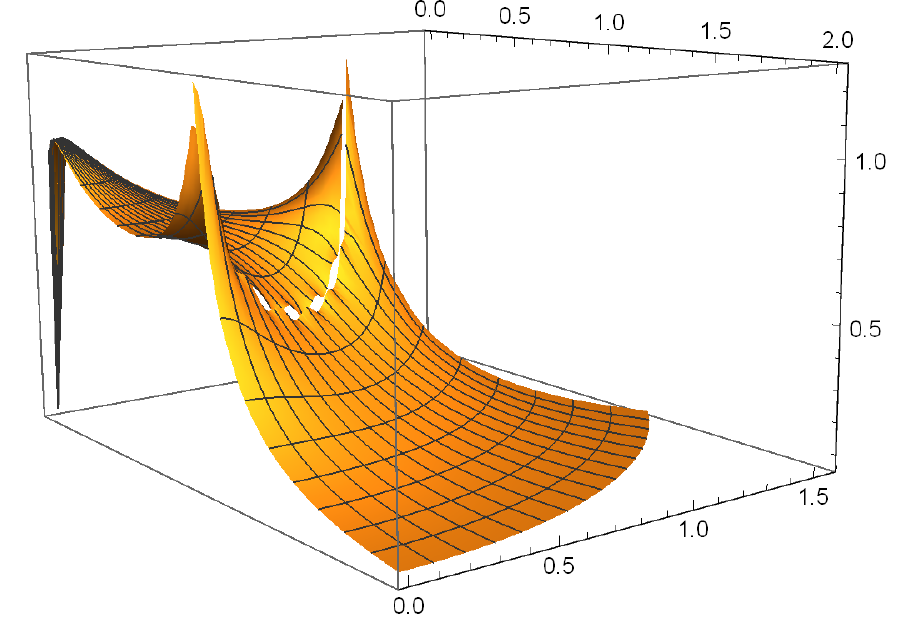}};
            \node at ( 4.5,4.5) [] {$ \operatorname{Re}\psi$};
            \node at ( 4.5,-0.1) [] {$ \operatorname{Im}\psi$};
            \node at ( 6.5,2.2) [] {$ g_{\psi, \bar{\psi}}$};
            \node at (7,0) [inner sep=0pt,above right] 
            {\includegraphics[width=0.45\textwidth]{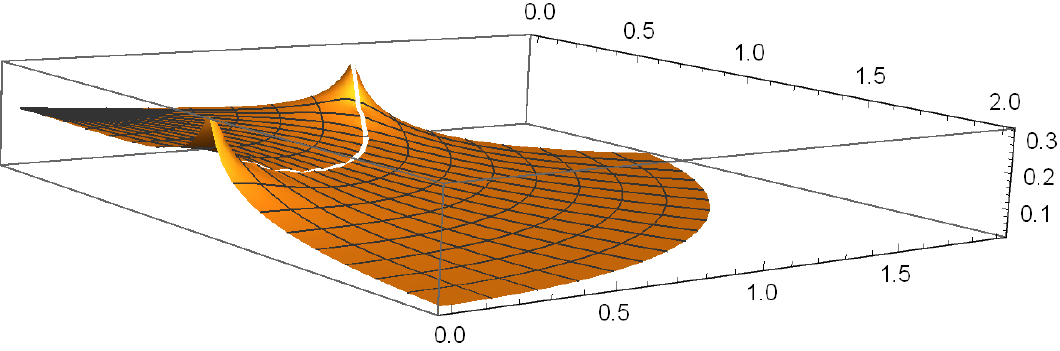}};
            \node at ( 12.2,2.3) [] {$ \operatorname{Re}\psi$};
            \node at ( 12,-0.2) [] {$ \operatorname{Im}\psi$};
            \node at ( 14.5,1.2) [] {$ g_{\psi, \bar{\psi}}$};
            \end{tikzpicture}
    \caption{Metrics for C1 (left) and the quintic (right).}
    \label{fig:metricVsPlotP6P4}
\end{figure}

\begin{figure}
        \centering
        \resizebox{0.4\textwidth}{!}
        {
            \begin{tikzpicture}
            \begin{axis}[
            legend style={
                legend pos=south east,},
            xlabel={$r$},
            ylabel={$g_{\psi \bar{\psi}}$},
            title =C1,
            ]
            \addplot [blue,] table [x index=0,y index=1] {p6phyMetric.dat};
            \addlegendentry{$\varphi = 0$};
            \addplot [red,dashed] table [x index=0,y index=2] {p6phyMetric.dat};
            \addlegendentry{$\varphi = \frac{\pi}{7}$};
            \addplot [orange,dashdotted] table [x index=0,y index=3] {p6phyMetric.dat};
            \addlegendentry{$\varphi = \frac{\pi}{14}$};
            \end{axis}
            \end{tikzpicture}
        }
        \caption{Metric for constant $\varphi$-values in the pseudo-hybrid phase of C1.}
        \label{fig:phyMetricC1}
\end{figure}
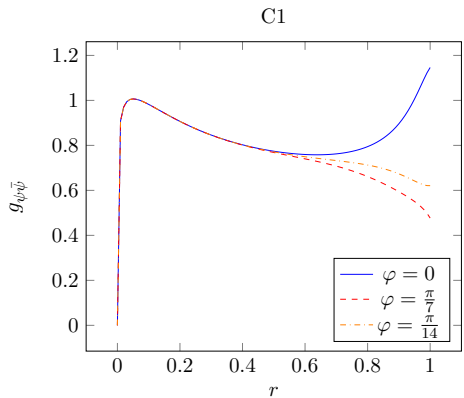

    In \cref{sec:ph1} we have computed the sphere partition function for this model. For the numerical calculations it is convenient to replace the variable $z$  by
    \begin{equation}
        z = - \frac{1}{2^4 3^3 \psi^7}.
    \end{equation}
     Furthermore, we write $\psi$ in terms of polar coordinates\footnote{We parameterize $\mathcal{M}_K$ using the ``classical'' K\"ahler parameter $z=e^{-t}$ with $t=2\pi\zeta-i\theta$. Since our focus is on computing geodesic distances, the results will be independent of the choice of parametrization. We choose the one that seems most natural for the GLSM point of view. If we were to extract Gromov-Witten invariants from $Z_{S^2}$, we would have to follow the steps outlined in \cite{Jockers:2012dk}.}:
   \begin{equation}
   \psi=r e^{i\varphi}=- \frac{1}{(2^4 3^3)^{1/7}} z^{-1/7}\qquad \Leftrightarrow \qquad r=\frac{1}{(2^4 3^3)^{1/7}}e^{\frac{2\pi \zeta}{7}}, \varphi=-\frac{\theta+ \pi}{7}.
   \end{equation}
   With this choice, the pseudo-hybrid point is at $\psi=0$ and the singular point at the phase boundary is at $(r,\varphi)=(1,0\,\mathrm{mod}\,\frac{2\pi}{7})$. Using this and the results \cref{eqn:phys2lr,eqn:phys2lg1,eqn:phys2lg2} of the sphere partition function, the leading behavior of the metric in the two phases is
    \begin{align}
      g_{\psi \bar{\psi}}^{\zeta \ll 0} &=  -\frac{2^8 7^3 \sqrt{3}  \pi^7  }{ 3^3 \Gamma \left(\frac{1}{6}\right)^4 \Gamma \left(\frac{1}{3}\right)^{10} }r^{1/3}\log(r)+\ldots , & g_{\psi \bar{\psi}}^{\zeta \gg 0} &= \frac{3}{4 r^2 \log(r)^2}+\ldots. \label{eqn:leading_ph1}
    \end{align}
    We plot the metric in \cref{fig:metricVsPlotP6P4}, along with the metric of the quintic.
As for the quintic, the limiting point at $\zeta\ll0$ is at finite distance in the moduli space, but the behavior is slightly different due to the logarithm: near the singular point the logarithm dominates and the behavior of the metric is similar to a large volume phase. Near the pseudo-hybrid point the polynomial behavior wins over so that the distance remains finite. This is shown in \cref{fig:phyMetricC1}. We only see a divergent behavior if the geodesic ends at the singular point at the phase boundary (solid line). From the plot one can also read off at which value of $r$ the dependence on the angular variable $\varphi$ sets in.

We are interested in computing geodesics that start near the pseudo-hybrid point and go all the way to the large volume phase. We call the distance inside the $\zeta\ll0$ phase $\Theta_0$ and calculate this distance as the
    geodesic distance to the phase boundary.  In order to get
    an approximation for the distance behavior in the $\zeta\gg0$ phase we 
    follow the discussion of \cite{Blumenhagen:2018nts}. Looking at the leading 
    behavior of the metric in $\zeta \gg 0$ phase and calculating the 
    distance $\Theta$ for a path with constant $\varphi$ one gets:
    \begin{equation}
        \Theta \approx \frac{1}{\lambda} \log(\abs{\log(\abs{z})}) + \frac{\alpha_1}{\log(\abs{z})^3} + \alpha_0 \label{eqn:larger_asym}.
    \end{equation}
    Replacing $z$ by $\psi$ gives:
    \begin{equation}
        \Theta \approx \frac{1}{\lambda} \log(\frac{1}{2 \pi}\log(2^4 3^3 r^7)) + \frac{\alpha_1}{\left(\frac{1}{2 \pi}\log(2^4 3^3 r^7)\right)^3} + \alpha_0
    \end{equation}
    We also define $\Theta_c$ as in (\ref{thetac}). We compute the geodesics for various starting values for $\varphi$. As a starting value for $r$ we choose\footnote{Due to numerical issues we could not start at $r=0$. The additional distance one should add by starting away from $0$ turns out to be negligible.} $ r = 10^{-6}$. Fitting against the asymptotic behavior, we get the values for the parameters that are summarized in \cref{tab:phyResults_ph1}. The angular variable $  \varphi$ takes values between $0 \leq \varphi \leq \frac{2 \pi}{7}$, but as one can see from the plot of the metric in \cref{fig:metricVsPlotP6P4}, there is a symmetry around $ \varphi = \frac{\pi}{7}$. For the fit we therefore focus on the region $0\leq\varphi\leq \frac{\pi}{7}$. Let us also comment on the  geodesics with small values of $\varphi$. These geodesics are  rather short in the large radius phase and they are not very useful for testing the conjecture. Nevertheless we kept them in our discussion, because we are mostly interested in the behavior in the small radius regime. There they display no behavior which would justify an exclusion. 
     The mean values of the fitted parameters are 
    \begin{align}
        \Theta_0 &\approx 0,8937,
         & \lambda^{-1} &\approx0,9608,
        & \Theta_c &\approx	1,8545.
         \label{eqn:fitvalues_phi1}
    \end{align}
    \begin{table}
        \centering
        \begin{tabular}{c|c|c|c|c|c}
            $\varphi(0)\frac{70}{\pi}$ & $\alpha_0$ & $\alpha_1$ & $ \lambda^{-1}$ & $\Theta_0$ & $\Theta_c$\\ \hline
             $1$	&	$1,0188$	&	$-0,066$	&	$0,6948$	&	$0,9213$	&	$1,6161$	 \\
             $2$	&	$0,8744$	&	$0,0734$	&	$1,1760$	&	$0,9155$	&	$2,0915$	 \\
             $3$	&	$0,8403$	&	$0,0951$	&	$1,1478$	&	$0,9067$	&	$2,0545$	 \\
             $4$	&	$0,8383$	&	$0,0847$	&	$0,9943$	&	$0,8982$	&	$1,8925$	 \\
             $5$	&	$0,8273$	&	$0,0898$	&	$1,0114$	&	$0,8924$	&	$1,9039$	 \\
             $6$	&	$0,8177$	&	$0,0935$	&	$0,9894$	&	$0,8872$	&	$1,8766$	 \\
             $7$	&	$0,8156$	&	$0,0890$	&	$0,9231$	&	$0,8822$	&	$1,8053$	 \\
             $8$	&	$0,8187$	&	$0,0834$	&	$0,9141$	&	$0,8796$	&	$1,7937$	 \\
             $9$	&	$0,8166$	&	$0,0831$	&	$0,8952$	&	$0,8777$	&	$1,7729$	 \\
             $10$	&	$0,8184$	&	$0,0792$	&	$0,8614$	&	$0,8764$	&	$1,7378$	 \\
        \end{tabular}
        \caption{Length parameters for C1.}
        \label{tab:phyResults_ph1}
    \end{table}
Our results are in agreement with the RSDC. Comparing to the quintic \cite{Blumenhagen:2018nts}, we see that the numerical values of $\Theta_0$ are by about a factor $2$ larger. 
\subsubsection{C2}
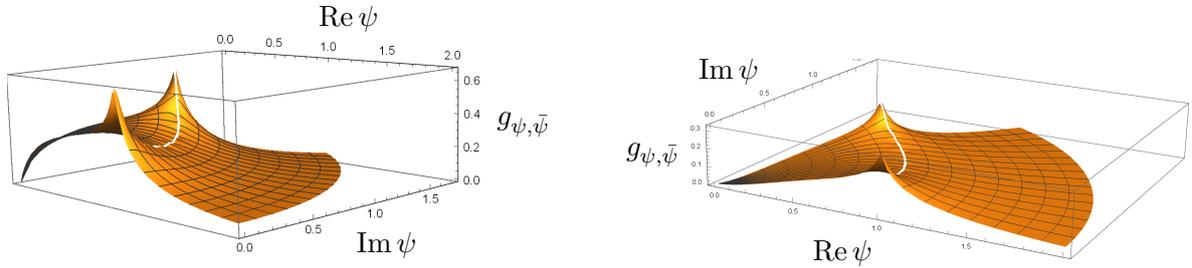
\begin{figure}
        \centering
        \begin{tikzpicture}
        \node [inner sep=0pt,above right] 
        {\includegraphics[width=0.4\textwidth]{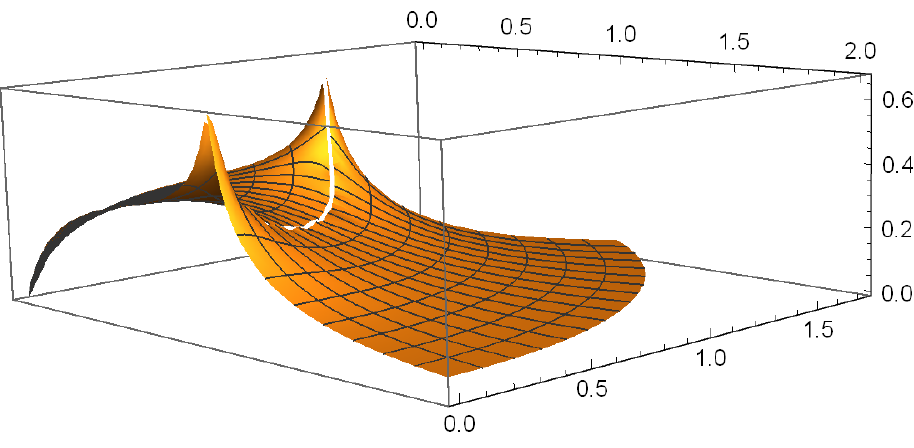}};
        \node at ( 4.5,3.2) [] {$ \operatorname{Re}\psi$};
        \node at ( 5,0.2) [] {$ \operatorname{Im}\psi$};
        \node at (7,0) [inner sep=0pt,above right] 
        {\includegraphics[width=0.7\textwidth]{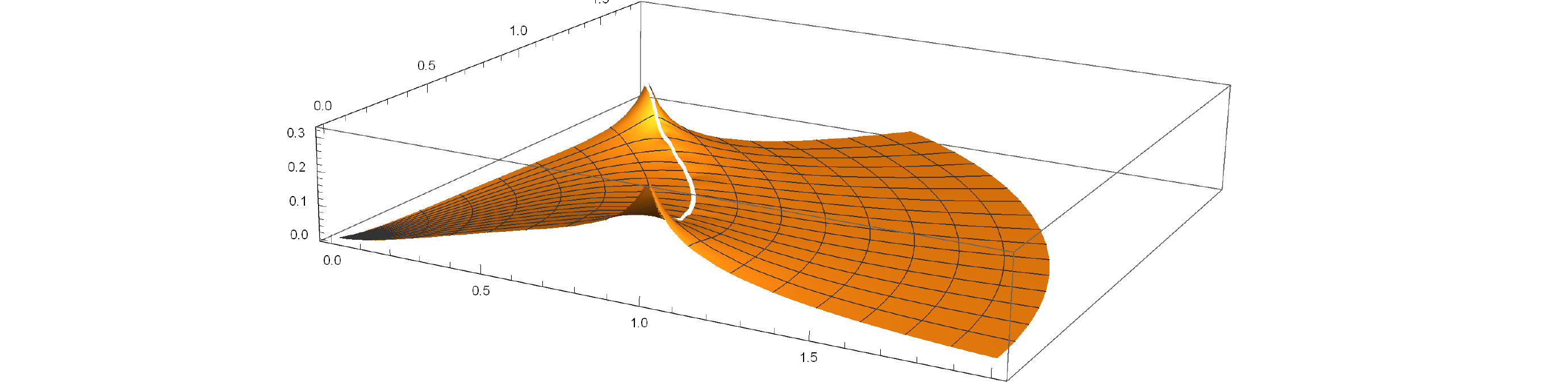}};
        \node at ( 9.5,2.5) [] {$ \operatorname{Im}\psi$};
        \node at ( 11,0.1) [] {$ \operatorname{Re}\psi$};
        \node at (8.5,1.4) [] {$ g_{\psi, \bar{\psi}}$};
        \node at ( 6.8,1.8) [] {$ g_{\psi, \bar{\psi}}$};
        \end{tikzpicture}
        \caption{Metrics for C2 (left) and C3 (right).}
        \label{fig:metricVsPlotP5}
\end{figure}

\begin{figure}
    \centering
    \resizebox{0.4\textwidth}{!}
    {
        \begin{tikzpicture}
        \begin{axis}[
        legend style={
            legend pos=south east,},
        xlabel={$r$},
        ylabel={$g_{\psi \bar{\psi}}$},
        title =C2,
        ]
        \addplot [blue,] table [x index=0,y index=1] {p5_4_2_phyMetric.dat};
        \addlegendentry{$\varphi = 0$};
        \addplot [red,dashed] table [x index=0,y index=2] {p5_4_2_phyMetric.dat};
        \addlegendentry{$\varphi = \frac{\pi}{6}$};
        \addplot [orange,dashdotted] table [x index=0,y index=3] {p5_4_2_phyMetric.dat};
        \addlegendentry{$\varphi = \frac{\pi}{12}$};
        \end{axis}
        \end{tikzpicture}
    }
    \resizebox{0.4\textwidth}{!}{
        \begin{tikzpicture}
        \begin{axis}[
        legend style={
            legend pos=south east,},
        xlabel={$r$},
        ylabel={$g_{\psi \bar{\psi}}$},
        title =C3,
        ]
        \addplot [blue,] table [x index=0,y index=1] {p5_6_2_phyMetric.dat};
        \addlegendentry{$\varphi = 0$};
        \addplot [red,dashed] table [x index=0,y index=2] {p5_6_2_phyMetric.dat};
        \addlegendentry{$\varphi = \frac{\pi}{6}$};
        \addplot [orange,dashdotted] table [x index=0,y index=3] {p5_6_2_phyMetric.dat};
        \addlegendentry{$\varphi = \frac{\pi}{12}$};
        \end{axis}
        \end{tikzpicture}
    }
    \caption{Metric for constant $\varphi$ values in the pseudo-hybrid phase for  C2 (left) and C3 (right).}
    \label{fig:phyMetricC2C3}
    \end{figure}
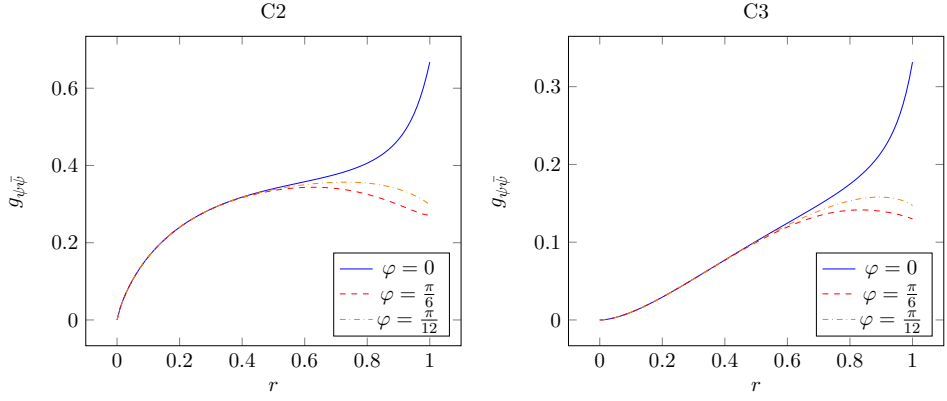

The analysis of this model is completely analogous to the first one. Using the results of the sphere partition function from \cref{sec:ph2}, we define
    \begin{equation}
        z = \frac{1}{1024 \psi^6},
    \end{equation}
   and further
    \begin{equation}
        \psi = r e^{i \varphi} = \frac{1}{(2^{10})^{1/6}}z^{-1/6} \quad 
       \Leftrightarrow \quad r = \frac{1}{(2^{10})^{1/6}}e^{\frac{2 \pi \zeta }{6}}, \varphi = -\frac{\theta}{6}.
    \end{equation}
    This parametrization moves the singularity to $(r,\varphi) = (1,0 \mod \frac{2 \pi}{6})$. The leading behavior of the metric in the two phases is 
    \begin{align}
        g_{\psi \bar{\psi}}^{\zeta \ll 0} &=-\frac{2^7 3^3 \pi ^6}{\Gamma \left(\frac{1}{4}\right)^{12}} r \log (r)+\ldots,
        & g_{\psi \bar{\psi}}^{\zeta \gg 0} &= \frac{3}{4 r^2 \log (r)^2}+\ldots. \label{eqn:leading_ph2}
    \end{align}
    A plot of the metric is given in \cref{fig:metricVsPlotP5}. 
The plot in \cref{fig:phyMetricC2C3} shows at which value of $r$ the angle dependence sets in.
The asymptotic behavior is the same as in (\ref{eqn:larger_asym}), and so we can write:
    \begin{equation}
    \Theta \approx \frac{1}{\lambda} \log(\frac{1}{2 \pi}\log(2^{10} r^6)) + \frac{\alpha_1}{\left(\frac{1}{2 \pi}\log(2^{10} r^6)\right)^3} + \alpha_0.
    \end{equation}
    A table with the fitted values for this model is given
    in \cref{tab:phyResults2}.
\begin{table}
        \centering
        \begin{tabular}{c|c|c|c|c|c}
            $\varphi(0)\frac{60}{\pi}$ & $\alpha_0$ & $\alpha_1$ & $ \lambda^{-1}$ & $\Theta_0$ & $\Theta_c$\\ \hline
        $1$	&	$0,5480$	&	$-0,0825$	&	$0,7328$	&	$0,5585$	&	$1,2913$	 \\
        $2$	&	$0,3417$	&	$0,1295$	&	$1,1551$	&	$0,5521$	&	$1,7073$	 \\
        $3$	&	$0,3706$	&	$0,1051$	&	$0,9735$	&	$0,5451$	&	$1,5186$	 \\
        $4$	&	$0,3441$	&	$0,1296$	&	$1,0043$	&	$0,5398$	&	$1,5440$	 \\
        $5$	&	$0,3327$	&	$0,1396$	&	$0,9918$	&	$0,5346$	&	$1,5265$	 \\
        $6$	&	$0,3376$	&	$0,1354$	&	$0,9309$	&	$0,5302$	&	$1,4610$	 \\
        $7$	&	$0,3321$	&	$0,1376$	&	$0,9369$	&	$0,5271$	&	$1,4640$	 \\
        $8$	&	$0,3391$	&	$0,1308$	&	$0,8924$	&	$0,5245$	&	$1,4169$	 \\
        $9$	&	$0,3328$	&	$0,1367$	&	$0,9001$	&	$0,5230$	&	$1,4231$	 \\
        $10$	&	$0,3323$	&	$0,1369$	&	$0,9015$	&	$0,5227$	&	$1,4242$	 \\
        \end{tabular}
        \caption{Length parameters for C2.}
        \label{tab:phyResults2}
    \end{table}
    Again we used the symmetry of the
    metric and focused on the region $\varphi \leq \frac{\pi}{6}$. Also in 
    this model we started from $ r = 10^{-6}$. The mean values of the fitted parameters are 
    \begin{align}
    \Theta_0 &\approx 0,5358,
    & \lambda^{-1} &\approx 0,9419,
    & \Theta_c &\approx	1,4777
    \label{eqn:fitvalues_ph2}.
    \end{align}
 This is again in agreement with the RSDC.   
    \subsubsection{C3}
    Using the results of \cref{sec:ph3} we define
    \begin{equation}
        z = \frac{1}{6912 \psi^6}
    \end{equation}
        and 
    \begin{equation}
        \psi = r e^{i \varphi} = \frac{1}{(2^8 3^3)^{1/6}} z^{-1/6} \quad  \Leftrightarrow \quad r = \frac{1}{(2^8 3^3)^{1/6}}e^{\frac{2 \pi \zeta}{6}}, \varphi = -\frac{\theta}{6}.
    \end{equation}
     This choice moves the singularity to
    $(r,\varphi) = (1,0 \mod \frac{2\pi}{6})$. In these coordinates 
    the leading behavior of the metric in the two phases is 
    \begin{align}
    g_{\psi \bar{\psi}}^{\zeta \ll 0} &=-\frac{2^5 3^5 \sqrt{3} \pi ^3\Gamma \left(\frac{5}{3}\right)^2 }{\Gamma \left(\frac{1}{6}\right)^8} r^2 \log(r) + \ldots, & g_{\psi \bar{\psi}}^{\zeta \gg 0} &= \frac{3}{4 r^2 \log (r)^2}+ \ldots.
    \label{eqn:leading_ph3}
    \end{align}
    The metric has been plotted in \cref{fig:metricVsPlotP5}.
    As in the previous two examples, we compute the lengths of geodesics in the $\zeta \ll 0 $ numerically,whereby as before we started from $ r = 10^{-6}$, and fit the geodesics in the $r\gg0$ phase against,
    \begin{equation}
    \Theta \approx \frac{1}{\lambda} \log(\frac{1}{2 \pi}\log(2^{8} 3^3 r^6)) + \frac{\alpha_1}{\left(\frac{1}{2 \pi}\log(2^{8} 3^3 r^6)\right)^3} + \alpha_0.
    \end{equation}
    The results are summarized in \cref{tab:phyResults3}.
    The mean values of the fitted parameters are 
    \begin{align}
    \Theta_0 &\approx 0,2832
    & \lambda^{-1} &\approx 0,9083
    & \Theta_c &\approx	1,1915, \label{eqn:fitvalues_ph3}
    \end{align}
which is in agreement with the RSDC.

    \begin{table}
        \centering
        \begin{tabular}{c|c|c|c|c|c}
            $\varphi(0)\frac{60}{\pi}$ & $\alpha_0$ & $\alpha_1$ & $ \lambda^{-1}$ & $\Theta_0$ & $\Theta_c$\\ \hline
          $1$	&	$0,1129$	&	$-0,1295$	&	$0,6783$	&	$0,2981$	&	$0,9764$	 \\
          $2$	&	$-0,1715$	&	$0,2836$	&	$1,0627$	&	$0,2936$	&	$1,3563$	 \\
          $3$	&	$-0,1032$	&	$0,2284$	&	$0,9076$	&	$0,2892$	&	$1,1968$	 \\
          $4$	&	$-0,1576$	&	$0,3030$	&	$0,9781$	&	$0,2858$	&	$1,2640$	 \\
          $5$	&	$-0,1486$	&	$0,3072$	&	$0,9370$	&	$0,2823$	&	$1,2193$	 \\
          $6$	&	$-0,1449$	&	$0,3112$	&	$0,9143$	&	$0,2797$	&	$1,1940$	 \\
          $7$	&	$-0,1530$	&	$0,3229$	&	$0,9194$	&	$0,2776$	&	$1,1970$	 \\
          $8$	&	$-0,1464$	&	$0,3295$	&	$0,8903$	&	$0,2760$	&	$1,1662$	 \\
          $9$	&	$-0,1583$	&	$0,3410$	&	$0,9105$	&	$0,2751$	&	$1,1857$	 \\
          $10$	&	$-0,1463$	&	$0,3314$	&	$0,8843$	&	$0,2747$	&	$1,1590$	 \\
        \end{tabular}
        \caption{Length parameters for C3.}
        \label{tab:phyResults3}
    \end{table}
    \subsubsection{Comparing C1, C2 and C3}
    Comparing the results for the distances $\Theta_0$ in the pseudo-hybrid phase (see \cref{eqn:fitvalues_phi1,eqn:fitvalues_ph2,eqn:fitvalues_ph3}), 
    we see that the values vary between the models. These differences can be
    explained by looking at the asymptotic behavior of the metric\footnote{This has also been computed \cite{Joshi:2019nzi}. In appendix \cref{sec:gammatrafo}
     we transform our results to the form given in \cite{Joshi:2019nzi}.}  
    \cref{eqn:leading_ph1,eqn:leading_ph2,eqn:leading_ph3}. 
    If we approach the point $r=0$ the $\log(r)$ contribution is strongly
    suppressed  by the polynomial behavior in the models C2 and C3. For
    the model C1 the logarithmic contribution is only suppressed by 
    $r^{1/3}$ and therefore we obtain the longest distance for these models.
    This behavior can also be seen in the plots of the metric \cref{fig:metricVsPlotP5,fig:metricVsPlotP6P4} and in more detail in the plots \cref{fig:phyMetricC1,fig:phyMetricC2C3}
    where we plotted the metric behavior in the pseudo-hybrid phase for 
    different values of $\varphi$. One sees that in the C1 model the logarithmic 
    behavior is much more dominant than for the other two models. 
    
    Comparing our results to the results given in \cite{Blumenhagen:2018nts}
    for the one parameter hypersurface examples we see that the C1 model shows
    the greatest deviations from the values given there. This can also be traced
    back to the different behavior of the metric in the pseudo-hybrid phase 
    compared to the behavior of the metric in the Landau-Ginzburg phase 
    of these models (see e.g. \cref{fig:metricVsPlotP6P4} for the quintic). Models with Landau-Ginzburg phases yield much smaller values for $\Theta_0$ than models with pseudo-hybrid phases. Among the pseudo-hybrid models, $\Theta_0$ grows larger the more the metric of the model deviates from the metric in a Landau-Ginzburg phase. The specifics of the phase are thus related to the values of the distance. Based on these considerations, it seems possible to make more precise statements on the bounds (\ref{refined}).

    The lengths associated to the large volume phases have a behavior comparable to the results of \cite{Blumenhagen:2018nts}. Whether the underlying geometry is a hypersurface or a complete intersection does not seem to have an obvious effect on the distance parameters. 
    \subsection{$F$-type example}
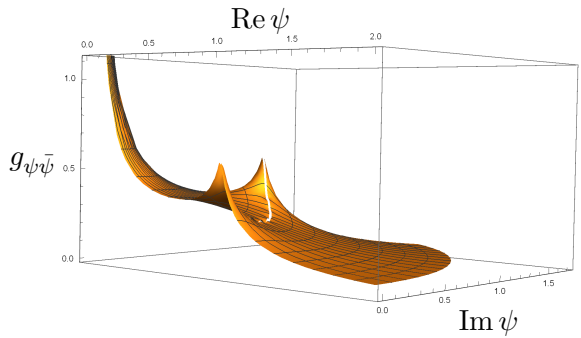
\begin{figure}
        \centering
        \begin{tikzpicture}
        \node [inner sep=0pt,above right] 
        {\includegraphics[width=1.0\textwidth]{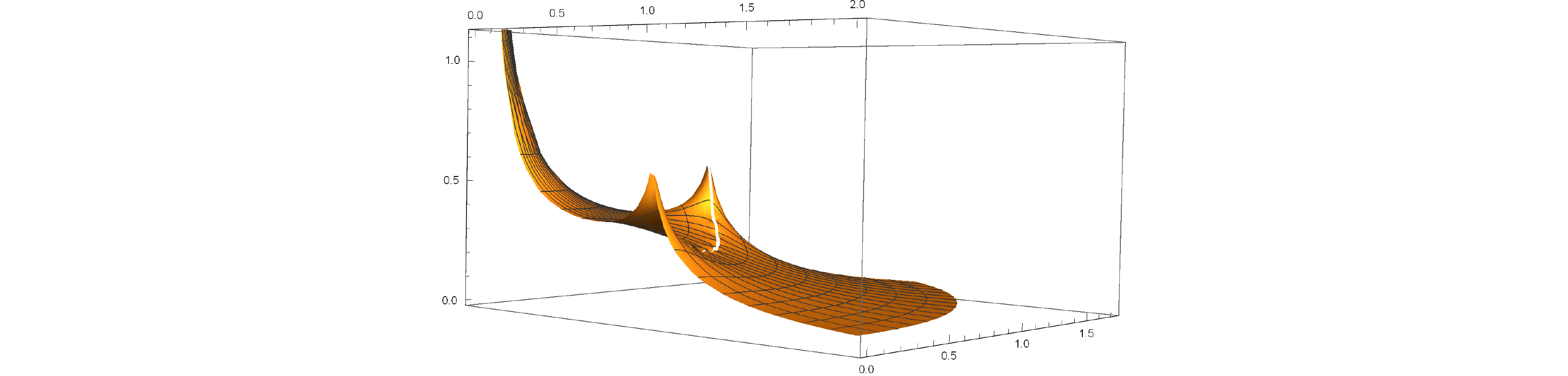}};
        \node at ( 7,3.9) [] {$ \operatorname{Re}\psi$};
        \node at ( 10,-0.1) [] {$ \operatorname{Im}\psi$};
        \node at (4,2) [] {$ g_{\psi\bar{\psi}}$};
        \end{tikzpicture}
        \caption{Metric plot of F1.}
        \label{fig:metricPlotF1}
    \end{figure}

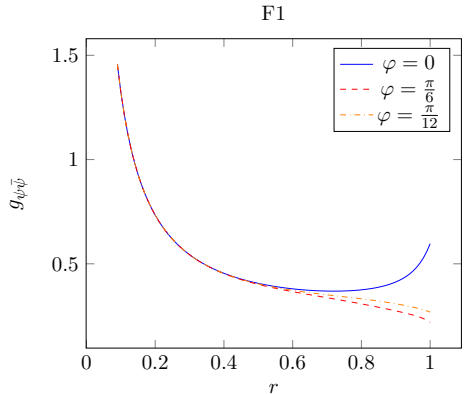
\begin{figure}
        \centering
        \resizebox{0.4\textwidth}{!}
        {
            \begin{tikzpicture}
            \begin{axis}[
            legend style={
                legend pos=north east,},
            xlabel={$r$},
            ylabel={$g_{\psi \bar{\psi}}$},
            title =F1,
            restrict y to domain=0:1.5,
            ]
            \addplot [blue] table [x index=0,y index=1] {p5_4_3_phyMetric.dat};
            \addlegendentry{$\varphi = 0$};
            \addplot [red,dashed] table [x index=0,y index=2] {p5_4_3_phyMetric.dat};
            \addlegendentry{$\varphi = \frac{\pi}{6}$};
            \addplot [orange,dashdotted] table [x index=0,y index=3] {p5_4_3_phyMetric.dat};
            \addlegendentry{$\varphi = \frac{\pi}{12}$};
            \end{axis}
            \end{tikzpicture}
        }
        \caption{Metric for constant $\varphi$-values in the pseudo-hybrid phase of F1.}
        \label{fig:phyMetricF1}
    \end{figure}

\begin{figure}
        \centering
        \resizebox{0.4\textwidth}{!}
        {
            \begin{tikzpicture}
            \begin{axis}[
            legend style={
                legend pos=north east,},
            xlabel={$r$},
            ylabel={$g_{\psi \bar{\psi}}$},
            title =Comparison of metrics,
            restrict y to domain=0:1.5,
            ]
            \addplot [blue] table [x index=0,y index=2] {p5_4_3_phyMetric.dat};
            \addlegendentry{F1};
            \addplot [purple,dashed] table [x index=0,y index=2] {p6phyMetric.dat};
            \addlegendentry{C1};
            \addplot [olive,dashdotted] table [x index=0,y index=2] {p5_4_2_phyMetric.dat};
            \addlegendentry{C2};
            \addplot [magenta,dashdotdotted] table [x index=0,y index=2] {p5_6_2_phyMetric.dat};
            \addlegendentry{C3};
            \end{axis}
            \end{tikzpicture}
        }
        \caption{Metric for the central $\varphi$-value in the $\zeta\ll0$-phases of C1, C2, C3 and F1.}
        \label{fig:phyMetricCentralComp}
    \end{figure}
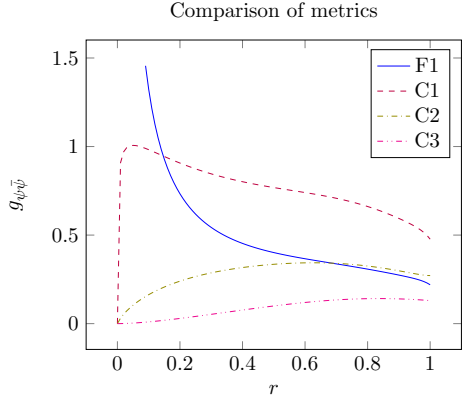

    The next example has been discussed in section \ref{sec-ftype}. We choose the coordinate $\psi$ such that
    \begin{equation}
      z=-\frac{1}{2^63^3\psi^6}.
    \end{equation}
    Given this, we define
    \begin{equation}
       r=\frac{1}{(2^6 3^3)^{1/6}}e^{\frac{2\pi \zeta}{6}}, \qquad\varphi=-\frac{\theta+ \pi}{6}.
    \end{equation}
    The remaining steps are the same as for the $C$-type model. We plot the metric in \cref{fig:metricPlotF1}. Even though $g_{\psi\bar{\psi}}$ is singular at $r=0$, the distance remains finite. This can be seen considering the leading order behavior of the metric in this phase:
    \begin{equation}
        g^{\zeta \ll 0}_{\psi \bar{\psi}} = \frac{3^3 \pi  \Gamma \left(\frac{1}{3}\right)^6 \Gamma \left(\frac{3}{4}\right)^2}{\Gamma \left(\frac{1}{4}\right)^{10}}\frac{1}{r} + \ldots.  \label{eqn:leading_f1}
    \end{equation}
    Integrating $\sqrt{ g^{\zeta \ll 0}_{\psi \bar{\psi}}}$ using (\ref{distanceint}) gives a finite contribution also for $r=0$.
    The leading behavior in the large radius phase is the same as in the other models.
     Plots of the metric for fixed values of $\varphi$ are depicted in \cref{fig:phyMetricF1}. 
    Due to the symmetry in $\varphi$ we consider geodesics for $0\leq\varphi\leq \frac{\pi}{6} $. Computing geodesics starting near\footnote{The starting value for the numerical calculation is $r=0,000001$. Using (\ref{distanceint}) one can compute the distance from $r=0$ to this value for a path with constant $\varphi$. One finds a contribution of $0,0007$ which we added as a correction to the numerical results for $\Theta_0$.} $r=0$ and ending in the $\zeta\gg0$-phase, we can numerically determine the length parameters as summarized in \cref{tab:resultsF1}. The mean values are 
    \begin{equation}
      \Theta_0 \approx 0,8286,
       \quad \Theta_\lambda \approx 0,9243,
       \quad \Theta_c \approx 1,7529.
    \end{equation}
        \begin{table}
        \centering
        \begin{tabular}{c|c|c|c|c|c}
            $\varphi(0)\frac{60}{\pi}$ & $\alpha_0$ & $\alpha_1$ & $ \lambda^{-1}$ & $\Theta_0$ & $\Theta_c$\\ \hline
            $1$	&	$0,7836$	&	$-0,1021$	&	$0,7365$	&	$0,8485$	&	$1,5850$	 \\
            $2$	&	$0,5787$	&	$0,1329$	&	$1,0728$	&	$0,8422$	&	$1,9150$	 \\
            $3$	&	$0,614$	&	$0,1051$	&	$0,9310$	&	$0,8366$	&	$1,7676$	 \\
            $4$	&	$0,5669$	&	$0,1538$	&	$1,0097$	&	$0,8322$	&	$1,8419$	 \\
            $5$	&	$0,5755$	&	$0,1503$	&	$0,9434$	&	$0,8275$	&	$1,7709$	 \\
            $6$	&	$0,5686$	&	$0,1567$	&	$0,9442$	&	$0,8243$	&	$1,7685$	 \\
            $7$	&	$0,5637$	&	$0,1627$	&	$0,9333$	&	$0,8212$	&	$1,7546$	 \\
            $8$	&	$0,5708$	&	$0,1588$	&	$0,8953$	&	$0,819$	&	$1,7143$	 \\
            $9$	&	$0,5601$	&	$0,1669$	&	$0,9226$	&	$0,8179$	&	$1,7405$	 \\
            $10$	&	$0,5788$	&	$0,1535$	&	$0,8542$	&	$0,8168$	&	$1,6711$	 \\
        \end{tabular}
        \caption{Length parameters for F1.}
        \label{tab:resultsF1}
    \end{table}
    We see that compared to the previous models, the distance in the $\zeta\ll0$-phase is larger than in the models C2 and C3 and larger than in the $F$-type models discussed in \cite{Blumenhagen:2018nts}. The corresponding distance in 
    the C1-model is slightly bigger. This behavior is as expected. This is illustrated in \cref{fig:phyMetricCentralComp},
    where we plotted the metric in the various model for the central $\varphi$ value. Though the metric in the F1-model
    diverges at the origin, it soon drops significantly under the value of the metric in the C1-model. In total, we again find agreement with the RSDC. 
        \subsection{Non-abelian example with a pseudo-hybrid phase}
Based on the results of section \ref{sec-nonab}, we can compute the distances $\Theta_0$ for geodesics that start at the pseudo-hybrid point and end at the $\zeta$-value of the nearest singular point. Given (\ref{nonab-sing}), we find
\begin{equation}
  \zeta_+=\frac{1}{2\pi}\log\frac{1}{540+312\sqrt{3}}\approx -6.99\qquad \zeta_-=\frac{1}{2\pi}\log\frac{1}{|540-312\sqrt{3}|}\approx 0.92.
\end{equation}
This means that $t_-$ is closer to the pseudo-hybrid phase. In order to simplify the calculation we introduce a coordinate $\psi$ in such a way that the pseudo-hybrid point is at $\psi=0$ and the nearest singular point is at $\psi=1$. Following \cite{Kanazawa:2012xya} it is convenient to choose
\begin{align}
z = (540-312 \sqrt{3}) \psi^7.
\end{align}
Switching to polar coordinates, we define
\begin{align}
\psi = r e^{i \varphi}, \qquad r = \frac{e^{-\frac{2 \pi}{7} \zeta}}{\abs{540-312 \sqrt{3}}^{\frac{1}{7}}} \quad \varphi = \frac{\theta + \pi}{7}.
\end{align} 
This moves the nearest singularity to $(r,\varphi) = (1,0 \mod 2 \pi)$.

The leading behavior of the metric is \begin{align}
g_{\psi \bar{\psi}}=-\frac{2^8 7^3 \left(2-\sqrt{3}\right) \pi ^8}{3^3 \Gamma \left(\frac{1}{6}\right)^2 \Gamma \left(\frac{1}{3}\right)^{14}}r^{\frac{1}{3}} \log (r) + \dots =-\alpha \frac{ 7^3 \Gamma \left(\frac{5}{6}\right)^9}{ 2^{\frac{8}{3}}\pi ^{\frac{9}{2}}}r^{\frac{1}{3}} \log (r) \dots,
\end{align}
with
\begin{equation}
\alpha =\left(\frac{2\sqrt{3}-3}{2^{\frac{2}{3}}}\right) \approx 0,292.
\end{equation}
We observe that the leading behavior is similar to the behavior of the  C1-example, except for the prefactor $\alpha$. This also gives us an additional check for the distances we compute numerically. Compared to C1, they should scale with a factor $\sqrt{\alpha}$. Note that $\alpha<1$, so the distances will be shorter than for C1.   

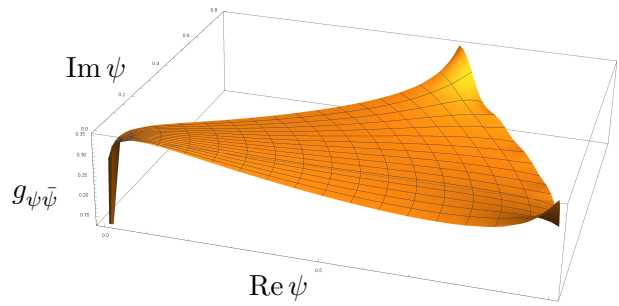
\begin{figure}
    \centering
    \begin{tikzpicture}
    \node [inner sep=0pt,above right] 
    {\includegraphics[width=0.5\textwidth]{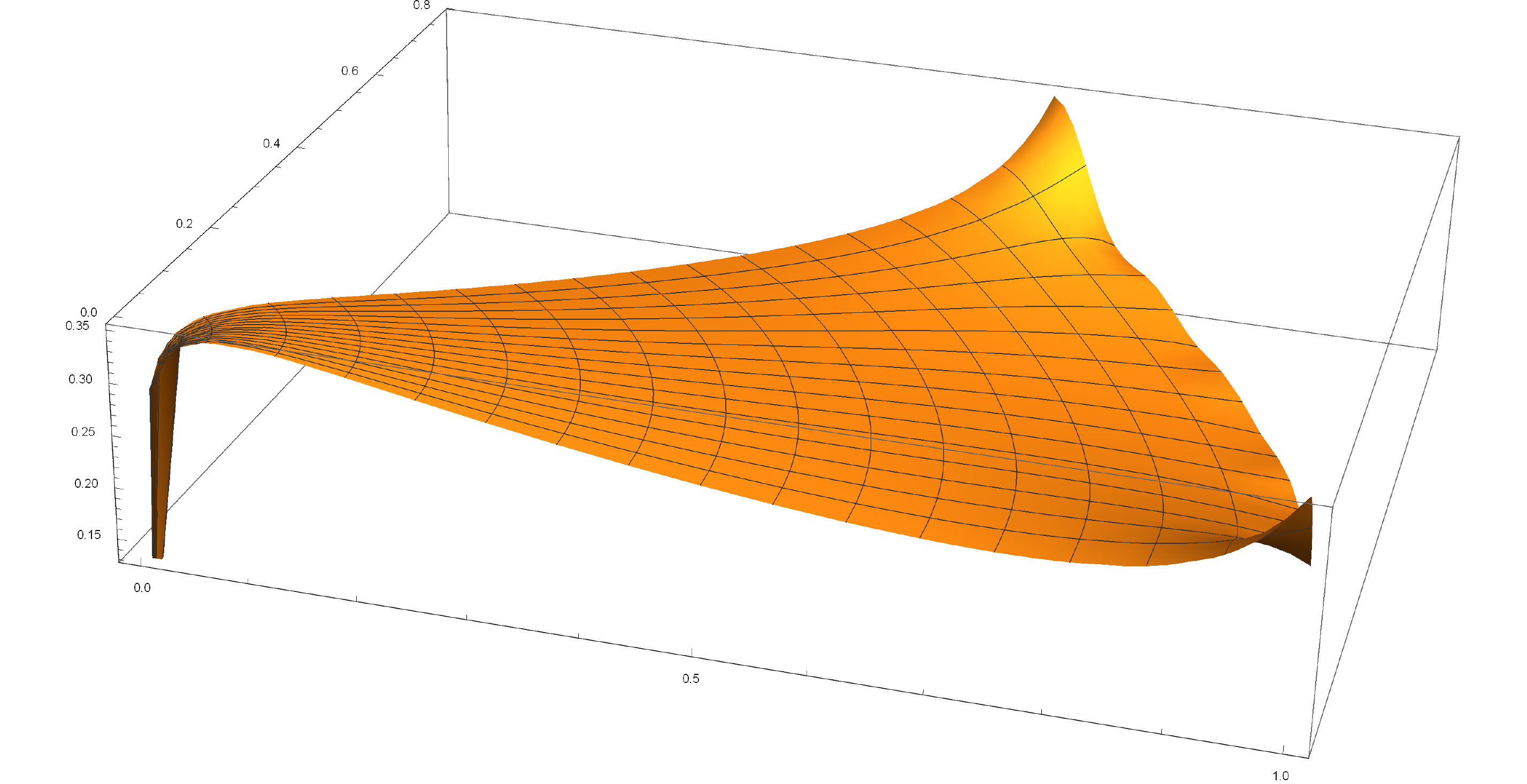}};
    \node at ( 3,0.3) [] {$ \operatorname{Re}\psi$};
    \node at ( 0.6,3.2) [] {$ \operatorname{Im}\psi$};
    \node at ( -0.2,1.5) [] {$ g_{\psi\bar{\psi}}$};
    \end{tikzpicture}
    \caption{Metric for the non-abelian model from the pseudo-point ($r=0$) to the nearest singularity ($r=1$).}
    \label{fig:X7metricFull}
\end{figure}

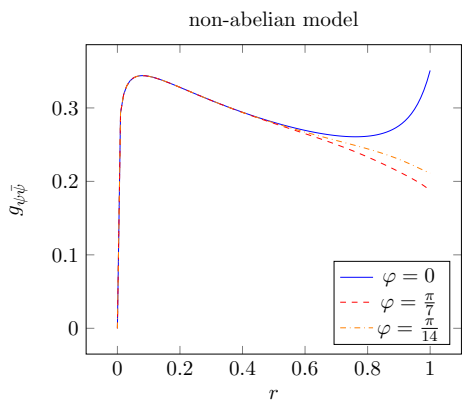
\begin{figure}
        \centering
    \resizebox{0.4\textwidth}{!}
    {
        \begin{tikzpicture}
        \begin{axis}[
        legend style={
            legend pos=south east,},
        xlabel={$r$},
        ylabel={$g_{\psi \bar{\psi}}$},
        title =non-abelian model,
        ]
        \addplot [blue,] table [x index=0,y index=1] {X7phyMetric.dat};
        \addlegendentry{$\varphi = 0$};
        \addplot [red,dashed] table [x index=0,y index=2] {X7phyMetric.dat};
        \addlegendentry{$\varphi = \frac{\pi}{7}$};
        \addplot [orange,dashdotted] table [x index=0,y index=3] {X7phyMetric.dat};
        \addlegendentry{$\varphi = \frac{\pi}{14}$};
        \end{axis}
        \end{tikzpicture}
    }
    \caption{Metric constant $\varphi$-values of the non-abelian model.}
    \label{fig:X7metric}
\end{figure}

The metric is plotted in \cref{fig:X7metricFull}.
As for the abelian models we observe a symmetry around $\varphi=\frac{\pi}{7}$. To see where the $\varphi$-dependence sets in, we plot the metric for specific values of $\varphi$ in \cref{fig:X7metric}.
The numerical results for the distances for various starting values of $\varphi$ are summarized in \cref{tab:phyResultsX7}.
\begin{table}
    \centering
    \begin{tabular}{c|c|c|c|c|c}
        $\varphi(0)\frac{70}{\pi}$ & 1&2 &3 &4 &5 \\ \hline
      $\Theta_{0} $ & $0,5406$ & $0,5376$ & $0,5343$ & $0,5319$ & $0,5298$  \\
      \hline\hline
    $\varphi(0)\frac{70}{\pi}$   &6&7 &7 &9 &10 \\ \hline
     $\Theta_{0} $  &$0,5277$ & $0,5263$ & $0,5253$	& $0,5246$ & $0,5243$
    \end{tabular}
    \caption{Length parameters for the non-abelian model.}
    \label{tab:phyResultsX7}
\end{table}
The mean value of the distance in the pseudo-hybrid phase is
\begin{equation}
  \label{th0nonab}
\Theta_{0} \approx 0,5303. 
\end{equation}
This is in agreement with the RSDC. Comparing to the C1-model, we find that
\begin{align}
\frac{\Theta_{0}}{\sqrt{\alpha}} \approx 0,8988.
\end{align}
This is in good agreement with $\Theta_{0}^{C1} \approx 0,8937$.

One can also get an approximation for $\Theta_0$ by computing the integral (\ref{distanceint}) for the leading term of the metric, which is independent of $\varphi$. In this case we find $\Theta_0\approx 0,4771$. This is significantly smaller than (\ref{th0nonab}). We can trace this back to the fact that we are integrating up to the boundary of the convergence radius at $r=1$ where the subleading terms give larger contributions\footnote{Of course the result is still convergent.}.  If we only integrate up to, say, $r=0.9$ the leading term is a good approximation. 
\section{Conclusions}
In this work we have shown that the RSDC holds for exotic hybrid CYs. We did this by explicitly computing the lengths of geodesics in the K\"ahler moduli space of one-parameter CY threefolds with pseudo-hybrid phases, and confirmed the RSDC. The specifics of the model were reflected in the numerical values for the distances. There are several interesting directions for further research.

One of them concerns CYs arising as phases of non-abelian GLSMs. While we have seen indications that the RSDC is satisfied for these models, the discussion is certainly not fully satisfactory and should be improved by addressing the challenges discussed in section \ref{sec-nonab}. We may return to this question in future work. In general, it might be worthwhile to study CYs that are not complete intersections in toric varieties in the context of swampland conjectures. A greater variety of examples might also improve the conjectures. For instance, one might be able to specify what ``$\mathcal{O}(1)$'' means more precisely. It would be interesting to study in more detail how the properties of the phases influence the numerical values of the lengths of the geodesics. A similar observation has been made in \cite{Blumenhagen:2018nts} where it was observed that the lengths are related to the number of moduli.

Another interesting question concerns the SDC for exotic limiting points such as hybrid points that are at infinite distance in the moduli space. For $K$-type hybrid models it has been conjectured in \cite{Joshi:2019nzi} that in infinite tower of D0 and D2-branes becomes massless. It would be interesting to understand this better, for instance with the help of the GLSM. 

Finally, it may be interesting to further study the pseudo-hybrid phases themselves in more detail. In this work we were mainly concerned with testing the RSDC and did not require the details about the low-energy description of the $C$- and $F$-points. Nevertheless, our rather superficial discussion revealed some curious properties. In order to understand this better, a discussion of the other $C$-type models along the lines of \cite{Aspinwall:2009qy} that also includes an analysis of the behavior of D-branes in these phases may be worthwhile. Even though they look simpler, a thorough analysis of the $F$-type examples may also be of interest. In this work we wrote down the most obvious abelian GLSMs for these models and obtained consistent results. As far as their construction in toric geometry is concerned, these models arise in a non-trivial way from models with more than one parameter \cite{Klemm:2004km,MR3409874}. It would be interesting to understand these connections better via the GLSM. 
\appendix
\section{Evaluating the sphere partition function}
\label{app-sphere}
Since we are only dealing with one-parameter models, the evaluation of the sphere partition function is relatively straight forward. See for instance \cite{Sharpe:2012ji} for a pedagogical discussion of a one-parameter example. We will discuss the details in a manner similar to \cite{Gerhardus:2015sla} that also works for higher dimensional residue integrals.

The general form 
of a contribution in the sphere partition function from a matter 
field is
\begin{equation}
  \label{typical}
    Z_\phi =\frac{\Gamma\left(a\tau +b\right)^\alpha}{\Gamma\left(c \tau +d\right)^\alpha}.
\end{equation}
The $\Gamma$-function has poles if the argument takes values in $ 
\mathbb{Z}_{\leq 0}$.  The expression (\ref{typical}) has poles in the numerator and the denominator:
\begin{align}
    a \tau + b &= -n, & c \tau +d &= - m
\end{align}
for $n,m \in \mathbb{Z}_{\geq 0}$.

The problem of the finding contributing poles can be recast into the geometric problem by introducing divisors. For a general contribution we get:
\begin{align}
    D^{n}_\phi &= a \tau +b + n, & D^{m}_\phi &= c \tau +d +m.
\end{align}

In the one-dimensional case, i.e. for $\mathrm{dim}\:\mathfrak{t}_{\mathbb{c}}=1$,  we can simply solve $D^{n}_\phi = 0$ to find the location of a pole:
\begin{equation}
    \tau_\phi^n = -\frac{n+b}{a}.
\end{equation}
We also have to take into account possible cancellations of a pole of the 
numerator by a pole of the denominator in (\ref{typical}). To derive the conditions for when this happens we insert $\tau_\phi^{n}$
into $D_\phi^{m} = 0$ and find
\begin{equation}
     m= \frac{c (n+b)}{a} - d.
\end{equation} 
Since a valid $m$ has to be in $\mathbb{Z}_{\geq 0}$, we find the 
condition
\begin{equation}
    \frac{c(n+b)}{a} \geq d,
\end{equation}
for a cancellation. As a consequence, the zeros of $D_\phi^n$ are only 
contributing poles if 
\begin{equation}
    d > \frac{c (n+b)}{a}.
\end{equation} 
\subsection{C1}
\label{sec:ap_pph1}
Let us now give the details of the evaluation of the sphere partition function of the pseudo-hybrid model discussed in section \ref{sec-pseudo}. Our starting point is (\ref{pseudo-z}).  The divisors for the pseudo-hybrid model, together with the conditions to avoid a cancellation are
\begin{align}
D_{p_1} &= 1+ 3 \tau + \frac{3}{2} m +n_{p_1} & n_{p_1} \geq \max \left[0,-3m\right], \label{eqn:abphydp}\\
D_{p_2} &= 1+2 \tau + m + n_{p_2} & n_{p_2} \geq \max\left[0,-2m\right],\label{eqn:abphydtildep} \\
D_x &= - \tau - \frac{1}{2} m + n_x & n_x \geq \max \left[0,m\right] 
\label{eqn:abphydphi},  
\end{align}
for $n \in \mathbb{Z}_{\geq 0}$. The poles lie along the real line, and we define the half-lines
\begin{align}
    H_1 &= \left\{\tau \in \mathbb{R} | \tau  > -q \right\}, & H_2 &= \left\{\tau \in \mathbb{R} | \tau  < -q \right\}. 
\end{align}
The exact location of the poles can be calculated from 
\cref{eqn:abphydp,eqn:abphydtildep,eqn:abphydphi} and results in
\begin{align}   
    \tau_{p_1} &= - \frac{1}{3}-\frac{1}{2}m - \frac{1}{3}n_{p_1}, \\
    \tau_{p_2} &= - \frac{1}{2} - \frac{1}{2} m - \frac{1}{2} n_{p_2}, \\
    \tau_x &= -\frac{1}{2}m + n_x.
\end{align}
For $0 < q < \frac{1}{3}$, one finds
\begin{align}
    \tau_x &\in H_1, &\tau_{p_1},\tau_{p_2} &\in H_2.
\end{align}
First, we consider the $\zeta\gg0$ phase, where the poles along $H_1$, i.e. those coming from $Z_{x}$, contribute. Introducing a shift $\tau \rightarrow \tau - \frac{1}{2} m +  n$, one can write all the contributions in terms of residues at $\tau=0$. The integrand of the sphere partition function becomes 
\begin{equation}
\begin{split}
Z_{p_1} &= \frac{\Gamma\left(1+3n+3\tau\right)}{\Gamma\left(3m-3n-3\tau\right)}, \quad Z_{p_2}=  \frac{\Gamma\left(1+2n+2\tau\right)^2}{\Gamma\left(2m-2n-2\tau\right)^2},\quad  Z_{x}= \frac{\Gamma\left(-n-\tau\right)^7}{\Gamma\left(1-m+n+\tau\right)^7}  \\
Z_{class} &= e^{-4 \pi \zeta \tau}e^{-4 \pi \zeta n}e^{(2\pi \zeta - i \theta)m} = \left(z \bar{z} \right)^{\tau+n} z^{-m},
\end{split}
\end{equation}
where we have introduced $z = ^{-2\pi \zeta + i \theta}$. To further simplify the calculation we define $a=n, b=n-m$. This simplifies the sums:
\begin{align}
    \sum_{n=0}^{\infty} \sum_{m \leq n } \rightarrow \sum_{a=0}^{\infty} \sum_{b=0}^{\infty}.
\end{align}
Further using the reflection formula $\Gamma(x)\Gamma(1-x)=\frac{\pi}{\sin\pi x}$, the terms in the integrand can be written as
\begin{equation}
\begin{split}
    Z_{p_1} &= \frac{\Gamma\left(1+3a+3\tau\right)}{\Gamma\left(-3b-3\tau\right)} =(-1)\frac{(-1)^{b}}{\pi} \sin\left(3\pi \tau\right)\Gamma\left(1+3a+3\tau\right)\Gamma\left(1+3b+3\tau\right)  \\
    Z_{p_2}&=  \frac{\Gamma\left(1+2a+2\tau\right)^2}{\Gamma\left(-2b-2\tau\right)^2} = \frac{\sin(2\tau \pi)^2}{\pi^2} \Gamma\left(1+2a+2\tau\right)^2\Gamma\left(1+2b+2\tau\right)^2\\
    Z_{x} &= \frac{\Gamma\left(-a-\tau\right)^7}{\Gamma\left(1+b+\tau\right)^7} =  (-1) (-1)^a \frac{\pi^7}{\sin(\pi \tau)^7} \frac{1}{\Gamma\left(1+b+\tau\right)^7\Gamma\left(1+a+\tau\right)^7} \\
    Z_{class} &= \left(z \bar{z} \right)^{\tau+n} z^{-m}= \left(z\bar{z}\right)^\tau \bar{z}^a z^b.
\end{split}
\end{equation}
Putting everything together, we arrive at the result in the main text.

Next, we consider the $\zeta\ll0$ phase, where the poles along $H_2$ contribute. Closing the contour to the left, two types of poles of $Z_{p_1}$ and $Z_{p_2}$ contribute. Due to cancellations of poles, we have to be mindful not to over-count. There are two possibilities, namely a pole of $Z_{p_1}$ is a simultaneous pole of $Z_{p_2}$ and vice versa. A careful analysis reveals whenever this happens the denominator of $Z_x$ cancels such a contribution.

Let us first focus on the contribution from $Z_{p_1}$. We shift the integration variable so that $ \tau \rightarrow \tau - \frac{1}{3} - \frac{1}{2} m - \frac{1}{3} n$. The contributions to the integrand become
\begin{equation}
\begin{split}
    Z_{p_1} &= \frac{\Gamma\left(-n+3\tau\right)}{\Gamma\left(1+3m+n-3\tau\right)}, \quad Z_{p_2}=  \frac{\Gamma\left(\frac{1-2n}{3}+2\tau\right)^2}{\Gamma\left(\frac{2+2n}{3}+2m -2\tau\right)^2}, \quad Z_{x} = \frac{\Gamma\left(\frac{1+n}{3}-\tau\right)^7}{\Gamma\left(\frac{2-n}{3}-m+\tau\right)^7}  \\
    Z_{class} &= e^{-4 \pi r \tau}e^{\frac{4}{3} \pi r (1+n)}e^{(2\pi r - i \theta)m} = \left(z \bar{z} \right)^{\tau} \left(z \bar{z} \right)^{- \frac{1+n}{3}} z^{-m}.
\end{split}
\end{equation}
To simplify the summations we introduce $l = 3m+n$ which gives the sums
\begin{equation}
    \sum_{n=0}^{\infty} \sum_{m \geq - \frac{n}{3}} \rightarrow  \left.\sum_{n=0}^{\infty} \sum_{l=0}^{\infty}\right|_{l-n \in 3\mathbb{Z}_{\geq 0}}.
\end{equation}
Note the constraint on the sums in $l$ and $n$ which is solved by
\begin{align}
    n &\in 3 \mathbb{Z}_{\geq 0}+\delta &&\rightarrow &l &\in 3 \mathbb{Z}_{\geq 0}+\delta\qquad \delta\in\{0,1,2\}.    
\end{align}
Hence, we introduce
\begin{equation}
    l= 3a +\delta \quad n = 3b + \delta, \quad m= a -b, \qquad \delta\in\{0,1,2\},
\end{equation}
so that we finally arrive at
\begin{equation}
    \left.\sum_{n=0}^{\infty} \sum_{l=0}^{\infty}\right|_{l-n \in 3\mathbb{Z}_{\geq 0}} \rightarrow \sum_{\delta =0}^{2}\sum_{a=0}^{\infty} \sum_{b=0}^{\infty}.
\end{equation}
The integrand of the sphere partition function becomes
\begin{equation}
\begin{split}
    Z_{p_1} &= \frac{\Gamma\left(-3b-\delta+3\tau\right)}{\Gamma\left(1+3a+\delta-3\tau\right)}, \quad Z_{p_2}=  \frac{\Gamma\left(\frac{1-2\delta}{3}-2b+2\tau\right)^2}{\Gamma\left(\frac{2+2\delta}{3}+2a -2\tau\right)^2}, \quad   Z_{x} = \frac{\Gamma\left(\frac{1+\delta}{3}+b-\tau\right)^7}{\Gamma\left(\frac{2-\delta}{3}-a+\tau\right)^7}  \\
    Z_{class} &= e^{-4 \pi \zeta \tau}e^{\frac{4}{3} \pi \zeta (1+n)}e^{(2\pi \zeta - i \theta)m} = \left(z \bar{z} \right)^{\tau} \left(z \bar{z}\right)^{- \frac{1+\delta}{3}} z^{-a} \bar{z}^{-b}.
\end{split}
\end{equation}
Applying the reflection formula as in the $\zeta\ll0$ phase, we arrive at the result in the main text.

Finally, we consider the contribution of $Z_{p_1}$. Defining $\tau \rightarrow \tau- \frac{1}{2} - \frac{1}{2} m - \frac{1}{2} n$, the integrand can be written as
\begin{equation}
\begin{split}
    Z_{p_1} &= \frac{\Gamma\left(- \frac{1+3n}{2}+3\tau\right)}{\Gamma\left(\frac{3+3n}{2}+3m -3\tau\right)},\quad   Z_{p_2}=  \frac{\Gamma\left(-n+2\tau\right)^2}{\Gamma\left(1+2m+n-2\tau\right)^2}, \quad Z_{x} = \frac{\Gamma\left(\frac{1+n}{2}-\tau\right)^7}{\Gamma\left(\frac{1-n}{2}-m+\tau\right)^7} \\
    Z_{class} &= e^{-4 \pi \zeta \tau}e^{2\pi \zeta (1+n)}e^{(2\pi \zeta - i \theta)m} = \left(z \bar{z} \right)^{\tau} \left(z \bar{z} \right)^{- \frac{1+n}{2}} z^{-m}.
\end{split}
\end{equation}
The sum in $m$ is restricted to $m \geq -\frac{n}{2}$. Defining $k = 2m+n$, the sums simplify to
\begin{equation}
    \sum_{n=0}^{\infty}\sum_{m \geq - \frac{m}{2}} \rightarrow \sum_{n=0}^{\infty} \sum_{k=0}^{\infty} \qquad \frac{k-n}{2} \in \mathbb{Z}.
\end{equation}
To remove the constraint on $k,n$ we further define
\begin{equation}
    n = 2 a + \delta, \quad k = 2 b + \delta, \quad m = b-a\qquad \delta\in\{0,1\},
\end{equation}
which yields
\begin{equation}
    \left. \sum_{n=0}^{\infty} \sum_{k=0}^{\infty} \right|_{k-n \in \mathbb{Z}}
    \rightarrow \sum_{\delta =0}^{1} \sum_{a=0}^{\infty} \sum_{b=0}^{\infty}.
\end{equation}
After these transformations the contributions to the integrand become
\begin{equation}
\begin{split}
    Z_{p_1} &= \frac{\Gamma\left(-\frac{1+3 \delta}{2} - 3 a + 3 \tau\right)}{\Gamma\left(\frac{3+3\delta}{2}+3b -3\tau\right)}, \quad Z_{p_2}=  \frac{\Gamma\left(-2a-\delta+2\tau\right)^2}{\Gamma\left(1+2b+\delta-2\tau\right)^2}, \quad Z_{x} = \frac{\Gamma\left(\frac{1+\delta}{2}+a-\tau\right)^7}{\Gamma\left(\frac{1-\delta}{2}-b+\tau\right)^7}  \\
    Z_{class} &=\left(z \bar{z} \right)^{\tau} \left(z \bar{z} \right)^{- \frac{1+\delta}{2}} z^{-b} \bar{z}^{-a}.
\end{split}
\end{equation}
Using the reflection formula the contribution to the sphere partition function becomes
\begin{align}
    Z_{S^2,Z_{p_2}}^{\zeta \ll 0} = \sum_{\delta =0}^{1} \left(z \bar{z}\right)^{{q}-\frac{1+\delta}{2}} \operatorname{Res}_{\tau=0} \left( \pi^{-4} \left(z \bar{z}\right)^\tau \frac{\sin \left(\pi \left(\frac{1-\delta}{2}-\tau\right)\right)^7}{\sin \left(\pi \left(-\frac{1+3\delta}{2}+3\tau\right)\right)\sin \left(2\tau \pi \right)^2}  f_2[\tau,z,\delta]\right),
\end{align}
with
\begin{equation}
    f_2[\tau,z,\delta] = \left|\sum_{a=0}^{\infty} (-z)^a \frac{\Gamma\left(\frac{1+\delta}{2}+a-\tau\right)^7}{\Gamma\left(
        1+2a+\delta-2\tau\right)^2\Gamma\left(\frac{3+3\delta}{2}+3a -3\tau\right)}\right|^2.
\end{equation}
Further investigation shows that only for $\delta = 0$ there is a second order pole, for $ \delta =1$ there is a cancellation of poles. Taking this into account, we finally arrive at the result in the main text. 

\subsection{C2}
\label{sec:ap_pph2}
Here we will give details on the evaluation of the sphere partition function for 
the model discussed in \cref{sec:ph2}. As most of the steps are similar to the 
evaluation described in \cref{sec:ap_pph1} we will only comment on the steps 
on how to avoid an over-counting of the poles. In this model 
an over-counting can only occur in the $\zeta \ll 0$ phase. This is obvious from the structure of the partition function (\ref{eqn:sppf_ph2}). 
Therefore we will only discuss this phase.

The contributions in the small radius phase come from $Z_{p_1}$  and $Z_{p_2}$.
Similar to the steps in \cref{sec:ap_pph1} we introduce the following 
divisors:
\begin{align}
D_{p_1} &= 2 m+n_1+4 \tau+1 & n_1 & \geq \max [0,-4m],\label{eqn:div1ph2} \\
D_{p_2} &=m+n_2+2 \tau+1 & n_2 & \geq [0,-2 m], \label{eqn:div2ph2} \\
D_{x} &= -\frac{m}{2}+n_3-\tau & n_3 & \geq [0,m].
\end{align}
The poles lie at 
\begin{align}
\tau^{Z_{p_1}} &=\frac{1}{4} \left(-2 m-n_{p_1}-1\right),\\
\tau^{Z_{p_2}} &=\frac{1}{2} \left(-m-n_{p_2}-1\right), \\
\tau^{Z_x} &=\frac{1}{2} \left(2 n_3-m_x\right) . 
\end{align}
Now one inserts a pole of $Z_{p_1}$ into (\ref{eqn:div2ph2}) and 
solves for $n_{p_2}$. This gives
\begin{align}
n_{p_2} &= \frac{n_1-1}{2}.
\end{align}
A valid $n_{p_2} $ must fulfill $n_{p_2} \in \mathbb{Z}_{\geq 0}$.
So it follows that poles coincide if:
\begin{align}
n_1 \in 2 \mathbb{Z}_{\geq 0} +1 \quad n_1 \geq 1.\label{eqn:pol1in2ph2}
\end{align}
Similarly, by inserting a 
pole of $Z_{p_2}$ into (\ref{eqn:div1ph2}) and solving for 
for $n_{p_1}$, one gets $n_1 = 2n_{p_2} +1$.
This is expected, considering (\ref{eqn:pol1in2ph2}). 
We know that $n_1 \in \mathbb{Z}_{\geq 0}$ and we see that 
every pole of $Z_{p_2}$ is a pole of $Z_{p_1}$. In order to avoid an over-counting we do the following. First we sum over all poles of $Z_{p_2}$ and get so all poles of $Z_{p_2}$ and the odd poles of $Z_{p_1}$. Then we sum over the even poles of $Z_{p_1}$and get the remaining poles of $Z_{p_1}$and avoid a double counting of poles of $Z_{p_2}$.

As for the evaluation in \cref{sec:ap_pph1} we will reduce the integral to a 
sum over residues. Let us start with the contribution from $Z_{p_2}$. 
We make the following transformation:
\begin{align}
\tau \rightarrow \tau- \frac{1}{2} - \frac{m}{2} - \frac{n_{p_2}}{2}.
\end{align}
In this case the sums read
\begin{align}
\sum_{n_{p_2} =0}^{\infty} \sum_{m\geq -\frac{n_{p_2}}{2}}.
\end{align}
To simplify the summation let us 
introduce $k = n_{p_2} +2m$. We get the constraint $k-n_{p_2} \in 2\mathbb{Z}_{\geq 0}$.
From this it follows that
\begin{align}
n_{p_2} \in 2 \mathbb{Z}_{\geq 0}+\delta  &\Rightarrow&  k \in 2 \mathbb{Z}_{\geq 0}+\delta \qquad \delta=\in\{0,1\}.
\end{align}
Therefore we introduce
\begin{align}
n_{p_2} = 2a + \delta \quad k = 2b + \delta.
\end{align} 
This yields the following sums
\begin{align}
\sum_{\delta=0}^{1}\sum_{a=0}^{\infty} \sum_{b=0}^{\infty}.
\end{align}
After all the transformations the integral contributions read:
\begin{align}
Z_{p_1} &= \frac{\Gamma \left(-4 a-2 \delta +4 \tau-1\right)}{\Gamma \left(4 b+2 \delta -4 \tau+2\right)} = - \frac{\pi}{\sin 4 \pi \tau} \frac{1}{\Gamma \left(4b+2 \delta -4 \tau +2\right)\Gamma \left(4a+2 \delta -4 \tau +2\right)},\\
Z_{p_2} &= \frac{\Gamma \left(-2 a-\delta +2 \tau\right)}{\Gamma \left(2 b+\delta -2 \tau+1\right)}
= (-1)^\delta\frac{\pi}{\sin 2 \pi \tau}  \frac{1}{\Gamma \left(2b+\delta +1 -2\tau\right)\Gamma \left(2a+\delta +1 -2 \tau\right)},\\
Z_x &= \frac{\Gamma \left(\frac{1}{2} \left(2 a+\delta -2 \tau+1\right)\right)}{\Gamma \left(-b-\frac{\delta }{2}+\tau+\frac{1}{2}\right)} \nonumber \\
& = - \frac{\left(\sin\left(\left(\frac{\delta-1}{2} -\tau\right) \pi\right)\right)}{\pi} (-1)^b \Gamma \left( a+ \frac{\delta+1}{2}-\tau\right) \Gamma \left( b+ \frac{\delta +1 }{2} -\tau\right),\\
Z_{class} &= e^{ \left(i \theta  (a-b)+2 \pi  \zeta (a+b+\delta -2 q+1)-4 \pi  \zeta \tau\right)}= (z\bar{z})^q (z \bar{z})^{- \frac{\delta +1}{2}} (z \bar{z})^{\tau} \bar{z}^{-a} z^{-b}.
\end{align}
Putting everything together we find the result in section \ref{sec:ph2}.

The contribution from $Z_{p_1}$ is a little more involved, because we have to consider a
restriction on the poles in order to avoid an over-counting. First we apply the transformation:
\begin{align}
\tau \rightarrow \tau - \frac{1}{4} - \frac{m}{2} - \frac{n_{p_1}}{4}.
\end{align}
Here we must restrict the sum to even $n_{p_1}$, because the odd values are already accounted for in $Z_{p_2}$. Therefore we make the replacement $n_{p_1} \rightarrow 2n$.
This results in the following sums
\begin{align}
\sum_{n=0}^{\infty} \sum_{m\geq -\frac{n}{2}} \rightarrow \sum_{n=0}^{\infty} \sum_{s=0}^{\infty},
\end{align}
where we introduced $s = 2n+4m$. But since $m \in \mathbb{Z}$ we must impose $s-2n \in 4\mathbb{Z}_{\geq0}$. In order to fulfill this condition we set
\begin{align}
s &= 4 a + \delta \qquad  2n = 4b + \delta\qquad \delta\in\{0,1,2,3\}
\end{align}
Due to $n \in \mathbb{Z}_{\geq 0}$ we see that only $\delta =0,2$ are allowed contributions. So we get the following sums
\begin{align}
\sum_{\delta = 0,2} \sum_{a=0}^{\infty}\sum_{b=0}^{\infty}.
\end{align}
After these transformations the various contributions read 
\begin{align}
Z_{p_1} &=\frac{\Gamma \left(-4 b-\delta +4 \tau\right)}{\Gamma \left(4 a+\delta -4 \tau+1\right)} = \pi\frac{ (-1)^{\delta}(-1)^{4b}}{\sin (4\pi \tau)}\frac{1}{\Gamma \left(1+4a +\delta -4\tau\right)\Gamma \left(1+4b+\delta -4\tau \right)}\\
Z_{p_2} &=\frac{\Gamma \left(-2 b-\frac{\delta }{2}+2 \tau+\frac{1}{2}\right)}{\Gamma \left(\frac{1}{2} \left(4 a+\delta -4 \tau+1\right)\right)} \nonumber \\
&= \pi \frac{(-1)^{2b}}{\sin \left(\pi \left(\frac{1-\delta}{2}+2\tau\right)\right)} \frac{1}{\Gamma \left(2a + \frac{1+\delta}{2} -2 \tau \right)\Gamma \left(2b + \frac{\delta +1}{2}-2 \tau\right)}\\
Z_x  &=\frac{\Gamma \left(b+\frac{\delta }{4}-\tau+\frac{1}{4}\right)}{\Gamma \left(-a-\frac{\delta }{4}+\tau+\frac{3}{4}\right)} \nonumber \\ &
= \pi^{-1}(-1)^a\sin \left(\pi \left(\frac{3-\delta}{4} +\tau\right) \right) \Gamma \left(b+ \frac{\delta +1}{4} -\tau\right)\Gamma \left(a + \frac{\delta +1}{4} -\tau\right)\\
Z_{class} &= e^{ \left(-i \theta  (a-b)+\pi  \zeta (2 a+2 b+\delta -4 q+1)-4 \pi  \zeta \tau\right)} = ( z \bar{z})^{q- \frac{\delta +1}{4}} (z \bar{z})^{\tau} z^{-a} \bar{z}^{-b}
\end{align}
Collecting all the results we get the expression in the main text.\\\\
We refrain from giving details on the evaluation of the sphere partition function in the model C3, because the discussion is completely analogous to C1 and C2. The same also holds for the model F1.
\subsection{Non-abelian model}
\label{app-nonab}
In this section we outline some details of the calculation of the multi-dimensional residue (\ref{znonab}) in the $\zeta\gg0$-phase of the non-abelian model. We follow the references \cite{russians,Friot:2011ic,Gerhardus:2015sla,Larsen:2017aqb}.

Finding the contributing poles in a multidimensional residue can be translated into a geometric problem of finding intersections of divisors associated to the poles of the integrand. We have already used this in the one-dimensional case, but only in the multi-dimensional case all aspects of the formalism become visible. For the non-abelian model we denote the divisors by $D_{k}^{n_k}$. These are defined by taking the arguments of the Gamma functions in the numerators of $Z_k$ and shifting them by an integer $n_k\in\mathbb{Z}_{\geq 0}$ that is constrained such that the corresponding Gamma function has a pole. For example,
\begin{equation}
  D_1^{n_1} =- \tau_1 - \tau_2 + \frac{1}{2} \left(m_1 +m_2 \right)+n_1  \quad n_1  \geq \max\left[0,-(m_1+m_2)\right].
\end{equation}
The $\zeta$-dependence enters into $Z_{\text{classical}}$. Convergence considerations divide the space $\tau\in\mathbb{R}^2$ into two half-spaces
\begin{align}
H_1&= \left\{ {\tau} \in \mathbb{R}^2 |  \tau_1 +\tau_2 > -2 {q} \right\} & H_2&= \left\{ {\tau} \in \mathbb{R}^2 |  \tau_1 +\tau_2 < -2 {q} \right\}. 
\end{align}
These are separated by the line
\begin{align}
\partial H  = \left\{ {\tau} \in \mathbb{R}^2 |  \tau_1 +\tau_2 = -2 {q}  \right\}.
\end{align}
Since we consider the case $\zeta\gg 0$, the poles in $H_2$ are relevant. Furthermore $\partial H$ gets divided into two half lines $\partial H_{\pm}$ by $\gamma$. This information can be used to define an orientation compared to the standard basis of $\mathbb{R}^2$. This will determine the signs in the residue. To determine the residues, one has to consider intersections of the divisors $D_{k}^{n_k}$ with $\partial H$. In our case, some of the divisors are parallel to $\partial H$ which is why $\partial H$ has to be slightly tilted. The intersection points can lie either on $\partial H_+$ or $\partial H_-$. The poles contributing to the $\zeta\ll0$-phase are those where pairs of divisors intersecting on $\partial H_+$ and $\partial H_-$, respectively, intersect in $H_2$. In our case, we have to consider the intersection of the divisors
\begin{equation}
  \{(D_3^{n_3},D_4^{n_4}),(D_3^{n_3},D_6^{n_6}),(D_5^{n_5},D_4^{n_4}),(D_5^{n_5},D_6^{n_6}) \}.
\end{equation}
The order is given in such a way that the orientation is always positive and all the residues come with a positive sign.

The next step is to exclude cancellations or double-counting of poles. This is quite tedious, but we have found that the following strategy works.
\begin{enumerate}
\item Contribution from $(D_3^{n_3},D_4^{n_4})$: Sum over $(n_3,n_4)\in\mathbb{Z}_{\geq 0}$.
\item Contribution from $(D_3^{n_3},D_6^{n_6})$: Sum over $(n_3,n_6)$ with the condition $n_3+2n_6\in 2\mathbb{Z}_{\geq0}$.
  \item Contribution from $(D_5^{n_5},D_4^{n_4})$: Sum over $(n_5,n_4)$ subject to the condition $n_5+2n_4\in 2\mathbb{Z}_{\geq0}$. 
\end{enumerate}
With this, the contribution from $(D_5^{n_5},D_6^{n_6})$ is automatically accounted for. Hence, we get three terms:
\begin{align}
Z_{S^2 }^{ \zeta \gg 0} = Z_{S^2,1 }^{ \zeta \gg 0} +Z_{S^2,2 }^{ \zeta \gg 0} +Z_{S^2,3 }^{ \zeta \gg 0} .
\end{align}
Due to symmetries it turns out that $Z_{S^2,3 }^{ \zeta \gg 0}=Z_{S^2,2 }^{ \zeta \gg 0}$. To evaluate the residues, we perform shifts of the summation variables that are very similar to the abelian case and arrive at similar expressions. For instance, we get
\begin{align}
\begin{split}
Z^{r \gg 0}_{S^2,1} = -\frac{1}{2}\sum_{\delta=0}^{2}& \left(z \bar{z}\right)^{-2q+ \frac{1+\delta}{3}}  \pi^{-3} \\ 
&\mathrm{Res}_{(x_1,x_2)=(0,0)}\left( \left(z \bar{z}\right)^{-x_1-x_2} \right. \\
&\left.\frac{ \sin \left(\pi  \left(\frac{2-\delta}{3} +x_1+ x_2\right)\right)^5 \sin \left( \pi  \left(\frac{1-2 \delta}{3} +2 x_1+2
   x_2\right)\right)^2\sin\left(\pi\left( \frac{2-\delta}{3}-x_2\right)\right)^3}{\sin \left(\pi  \left(2 x_1+x_2\right)\right)^2\sin \left(\pi  \left(x_1+2 x_2\right)\right)^2\sin \left(\pi\left(\frac{1-2\delta}{3}+x_1\right)\right)^3} \right. \\
&\left. \left| f_1[z,x_1,x_2,\delta] \right|^2\right),
\end{split}
\end{align}
with $f_1$ defined in the main text. The explicit computation of the residue is problematic due the $\sin$-functions in the denominator. To evaluate this expression we use the following property of the Grothendieck residue at $x=0$:
\begin{align}
\mathrm{Res}_{x=0} \left( \frac{\omega(x)dx_1\wedge \dots \wedge dx_n}{f_1(x)\dots f_n(x)}\right) =\mathrm{Res}_{x=0} \left( \frac{\mathrm{Det}(A(x))\omega(x)dx_1\wedge \dots \wedge dx_n}{g_1(x)\dots g_n(x)}\right),
\end{align}
with $g_i=\sum_j A_{ij}f_j$. Applying the identity
\begin{align}
(a+b)^{2N-1} = b^N \sum_{k=0}^{N-1}
 \begin{pmatrix}
2N-1 \\
k
\end{pmatrix}
a^kb^{N-1-k}
+
a^N \sum_{k=0}^{N-1}
 \begin{pmatrix}
2N-1 \\
k+N
\end{pmatrix}
a^kb^{N-1-k},
\end{align}
we can rewrite bring the expression into a form $Z^{r \gg 0}_{S^2,1}=\mathrm{Res}_{x=0}\frac{\mathrm{Det}(A(x))h(x)}{x_1^kx_2^l}$, where $h(x)$ is regular. After further modifications and repeated use of the reflection formula we can argue that $f_1[z,0,0,\delta]=0$ and
\begin{align}
\partial_{x_1} f_1[z,x_1,x_2,\delta]|_{(0,0)} = -\partial_{x_2} f_1[z,x_1,x_2,\delta]|_{(0,0)}.
\end{align}
Furthermore we find
\begin{align}
\partial_{x_1} (fg)|_{(0,0)} &= 0, \\
\partial_{x_1}\partial_{x_2} (fg)|_{(0,0)} &=  \partial_{x_1}f\partial_{x_2} g +   \partial_{x_2}f\partial_{x_1} g = -2 \partial_{x_1}f\partial_{x_1}g = - \partial^2_{x_1} (fg)|_{(0,0}, \\
\partial^2_{x_1} (fg)|_{(0,0)} &= \partial^2_{x_2}(fg)|_{(0,0)},
\end{align}
with $f=f_1[z,x_1,x_2,\delta]$ and $g=f_1[\bar{z},x_1,x_2,\delta]$. Using this we arrive at the result in section \ref{sec-nonab}. The evaluation of $Z_{S^2,2 }^{ \zeta \gg 0}$ goes along similar lines. 
\section{Transforming the leading metric behavior}
\label{sec:gammatrafo}
 Here we will show the necessary transformations to transform the leading behavior of the metric in the pseudo hybrid phase to the form given in \cite{Joshi:2019nzi}.
We will use the following $\Gamma$-function identities:
\begin{align}
\Gamma(1+z) &= z \Gamma(z), \\
\Gamma(z)\Gamma(1-z) &= \frac{\pi}{\sin \pi z}. \label{eqn:gammareflexion}
\end{align}
Furthermore we use
\begin{align}
\Gamma\left(\frac{1}{6}\right) &= 2^{-\frac{1}{3}} 3^{\frac{1}{2}} \pi^{-\frac{1}{2}} \Gamma \left(\frac{1}{3}\right)^2.
\label{eqn:gamma16}
\end{align}

\subsection{C1}
Here we use $\Gamma\left(- \frac{2}{3} \right) = - \frac{3}{2} \Gamma\left(\frac{1}{3}\right)$, (\ref{eqn:gamma16}), and (\ref{eqn:gammareflexion}) for $z=\frac{1}{6}$. Given this, one can show that (\ref{eqn:leading_ph1}) can be written in the following form:
\begin{align}
g_{\psi \bar{\psi}}^{\zeta \ll 0} &=  -\frac{2^8 7^3 \sqrt{3}  \pi^7  }{ 3^3 \Gamma \left(\frac{1}{6}\right)^4 \Gamma \left(\frac{1}{3}\right)^{10} }r^{1/3}\log(r) 
=  -\frac{7^3 \Gamma\left(\frac{5}{6}\right)^9  }{ 2^{\frac{8}{3}}\pi^{\frac{9}{2}} }r^{1/3}\log(r).
\end{align}

\subsection{C2}
By   (\ref{eqn:gammareflexion}) we can write (\ref{eqn:leading_ph2}) as
\begin{align}
g_{\psi \bar{\psi}}^{\zeta \ll 0} &=-\frac{2^7 3^3 \pi ^6}{\Gamma \left(\frac{1}{4}\right)^{12}} r \log (r) 
= -\frac{2 3^3 \Gamma\left(\frac{3}{4}\right)^{12}}{\pi^6} r \log(r).
\end{align}

\subsection{C3}
We use the following identity:
\begin{align}
\Gamma \left(\frac{5}{3}\right) &= \Gamma \left(1+ \frac{2}{3}\right) =\frac{2}{3} \Gamma\left(\frac{2}{3}\right) 
= \frac{2}{3} \frac{2 \pi}{\sqrt{3} \Gamma\left(\frac{1}{3}\right)},
\end{align}
and apply (\ref{eqn:gamma16}). Further using (\ref{eqn:gammareflexion}) on $\Gamma\left(\frac{1}{6}\right)$ we can show that (\ref{eqn:leading_ph3}) can be transformed to:
\begin{align}
g_{\psi \bar{\psi}}^{\zeta \ll 0} &=-\frac{2^5 3^5 \sqrt{3} \pi ^3\Gamma \left(\frac{5}{3}\right)^2 }{\Gamma \left(\frac{1}{6}\right)^8} r^2 \log(r)
= - \frac{3^3\Gamma \left(\frac{5}{6}\right)^9 }{2^{\frac{1}{3}}\pi^{\frac{9}{2}}} r^2 \log(r).
\end{align}

\subsection{F1}
Here we simply use (\ref{eqn:gammareflexion}) with $z=\frac{1}{4}$. It follows that (\ref{eqn:leading_f1}) can be rewritten as
\begin{align}
g^{\zeta \ll 0}_{\psi \bar{\psi}} &= \frac{3^3 \pi  \Gamma \left(\frac{1}{3}\right)^6 \Gamma \left(\frac{3}{4}\right)^2}{\Gamma \left(\frac{1}{4}\right)^{10}}\frac{1}{r} 
= \frac{3^3   \Gamma \left(\frac{1}{3}\right)^6 \Gamma \left(\frac{3}{4}\right)^4}{2 \pi\Gamma \left(\frac{1}{4}\right)^{8}}\frac{1}{r}.
\end{align}


\bibliographystyle{fullsort}
\bibliography{distance}
\end{document}